\documentclass[twocolumn]{revtex4-2}

\usepackage[utf8]{inputenc}
\usepackage{babel}
\usepackage{fontenc}
\usepackage{graphicx}
\usepackage{amsmath}
\usepackage{subcaption}
\usepackage{verbatim}
\usepackage{color}
\usepackage{amssymb}

\usepackage[normalem]{ulem}

\begin{document}

\author{Dalibor Jav\r{u}rek}
\email{dalibor.javurek@mojeposta.xyz}
\affiliation{Department of Physical Electronics, 
Faculty of Nuclear Sciences and Physical Engineering,
Czech Technical University in Prague,
B\v{r}ehov\'{a} 7, 11519 Prague 1,
Czech Republic}

\date{\today}

\title{Multiple measurements on an uncollapsed entangled two-photon state}

\begin{abstract}
 The relativity of simultaneity together with the definition of a quan\-tum state collapse \cite{Susskind2014quantum,Griffiths2004introduction,Wechsler2021quantum} result into experimental situations, where multiple measurements can be taken on an uncollapsed quantum state. The quantum state's collapse is defined to be instantaneous in a rest inertial frame of a detector performing measurements on the quantum system. The definition is consistent with Copenhagen interpretation and in agreement with all measurements performed with detectors at rest in arbitrary Lorentz (laboratory) frame. From the introduced collapse model follows, that under certain conditions, multiple measurements are allowed on the same uncollapsed quantum state. An application of the developed approach is shown on measurement of photon-pair state entangled in polarization and energy. Conditions, under which two mea\-sure\-ments can be taken on the uncollapsed photon-pair state, are derived. Serious consequences follow from allowance of multiple measurements on the same uncollapsed state. For example, the mea\-sure\-ments taken by both detectors in this situation are uncorrelated. Moreover, all the conservation laws could be violated in individual measurements, but not in mean values. This statement is proved on the two-photon state entangled in energy. This is in contradiction with experimental results observed by the detectors in rest relative to each other. It is shown, that the property of measuring uncorrelated results with detectors in relative movement is related solely to the proposed collapse model. The remaining collapse models -- Preferred Lorentz frame, Aharonov-Albert and Hellwig-Kraus are examined and discussed with respect to the designed experiment, which involves space-like separated measurements.
\end{abstract}

\maketitle

 \section{Introduction}
 A state of a quantum system undergoes two types of space-time evolution \cite{Susskind2014quantum,Griffiths2004introduction,Ohanian2017}. When the quantum system is undisturbed by a measurement apparatus (detector), time evolution of its quantum state (QS) is given by the Schr\"odinger equation. Since the time evolution of the quantum system can be uniquely predicted by solution of the equation, the time evolution is considered to be deterministic. When the quantum system interacts with the measurement apparatus, its QS undergoes a collapse. The collapse is explained as projection of the quantum system's QS into eigen-vector related to a value measured by the detector. This process is considered to be purely statistical, since only the probabilities with which the values can be measured (on the quantum system) can be predicted. The QS's collapse, according to the Copenhagen interpretation, occurs at the time of the measurement on all space \cite{Susskind2014quantum,Ohanian2017}.

 First discrepancies between relativity of simultaneity and time evolution of a QS have been described by Bloch \cite{Bloch1967some}. He studied interaction of a QS with multiple detectors positioned at various points in space-time when viewed (and evolved) in different Lorentz frames. He concluded, that ambiguities in QS may arise, when it is evolved in moving Lorentz frames. But these ambiguities do not affect probabilities of the measured results. He also proposed the Lorentz-invariant QS's collapse along the backward light cone. This idea was later developed by Hellwig and Kraus \cite{Hellwig1970formal}. Aharonov and Albert argued the preferred reference frame models and the Hellwig-Kraus model of wave-function's collapse are un\-sui\-ta\-ble \cite{Aharonov1981can,Cohen1995retrodiction}. Instead, they proposed QS's collapse to occur instantaneously for any observer. They based their argumentation on measurement of the nonlocal observables. All models of the state's reduction have been summarized and discussed with respect to three most used in\-ter\-pre\-ta\-tions of QM in review \cite{Cohen1995retrodiction} by Cohen and Hiley. The authors claim to defend the preferred Lorentz frame (PLF) model and extended it to include unitary interactions. The PLF model defines the collapse of the QS to occur instantaneously in a preferred (particular) Lorentz frame. In their work C\&H~\cite{Cohen1995retrodiction} were exploring space-time properties of a general PLF model utilizing single collapse hypersurface.

 Cohen and Hiley (C\&H) showed \cite{Cohen1995retrodiction} Aharonov's and Albert's (A\&A) arguments for discarding of the PLF model to be invalid. They managed to prove that both models provide the same results of a measurement. C\&H studied measurement of nonlocal observables, as proposed by A\&A, on singlet spin state $\vert \Psi \rangle = \vert +_z \rangle_1 \vert -_z \rangle_2 - \vert -_z \rangle_1 \vert +_z \rangle_2$ ($-_z$ denotes spin down projected on axis $z$ and $+_z$ spin up in the same direction). A\&A showed, that utilization of immediate collapse in the measurement (on QS  $\vert \Psi \rangle$) results in non-demolishing verification of the state. On the other hand, they argued that collapse occurring in preferred Lorentz frame leads to different measurement results~\cite{Cohen1995retrodiction}. Cohen and Hiley (C\&H) managed to disproved this by an explicit calculation of the final state, which emerges after the measurement.

C\&H further argue that A\&A collapse model does not propose unique space-time hypersurface for collapse of a QS. Because of this property, C\&H seem to prefer the PLF model. It allows them to introduce unique unitary evolution of a QS including interactions. A\&A are not concerned about ambiguities of the QS's collapse or physical meaning of the QS as far as it provides correct results of the measurements. We note, that this treatment may appear to be incorrect if Susskind's and Maldacena's recent hypothesis in its strong form reveals to true \cite{Maldacena2013cool,Susskind2016copenhagen}. It states, that entanglement between quantum particles is physically realized by Einstein-Rosen bridges (i.e. wormholes in space-time). Then, a QS of a quantum system should be treated as an object of reality with unique properties as well. We remark, that analysis of closing of the exits of the wormholes by the external nearby moving observers may bring a significant shift in both topics -- Susskind's and Madacena's hypothesis as well as relativity of the QS's collapse.
 
Hellwig and Kraus (H\&K) model requires collapse of a QS to occur before a measurement takes place \cite{Cohen1995retrodiction,Ohanian2017}. It is the only model, which preserves the collapse hypersurface under the Lorentz transformation. The property of collapsing of the QS prior to the measurement is as well called a pre-collapse \cite{Ohanian2017}. This property is in contradiction with causality. However, there has been no argument or experiment revealing that the violation of causality can be demonstrated by a measurement. Similarly to A\&A model, H\&K model posses QS's ambiguities (see section~9 in~\cite{Cohen1995retrodiction}). The ambiguities have their roots in elements of the H\&K model.  Because of these properties, H\&K model is considered as inappropriate \cite{Cohen1995retrodiction}. In Appendix~B we show, for the first time, that Hellwig-Kraus model predicts results conflicting with already performed experiments. This is observed, when two space-like separated measurements are performed on an entangled state.

The Preferred Lorentz frame (PLF) collapse model defines a collapse of a quantum state to occur instantaneously in only one chosen Lorentz frame (called preferred frame). This unique frame may not necessarily be attached to any detector nor observer. In consequence, all collapse hypersurfaces related to measurements by the detectors will be parallel to each other. The preferred frame could be attached to center of mass frame of two-particle state or to microwave background radiation frame~\cite{Cohen1995retrodiction}. The PLF model has already been used for studying collapse properties of an entangled electron-pair \cite{Cohen1995retrodiction,Aharonov1981can}.

In this paper, we introduce Multiple Lorentz frame collapse model (MLF), which sets the collapse of a QS to occur immediately in a rest frame of a detector taking mea\-sure\-ment on the QS. Since there can be more detectors moving relative to each other and taking measurements on the QS, the collapse hypersurfaces related to individual measurements may intersect. Due to this property, the area where the QS is uncollapsed and collapsed after series of measurements has to be properly identified.

A motivation for development of the MLF collapse model is to generalize Copenhagen interpretation (CI) of collapse of the QS to measurement schemes, where the detectors are in relative movement to each other (relativistic schemes). The features of the MLF collapse model are: 1. It attaches unique collapse hypersurface to a measurement and 2. Preserves compatibility with CI in situations, when all detectors are at rest in one laboratory Lorentz frame. These properties makes the MLF model unique among the others. In all experiments performed so far, the detectors have been at rest in one (laboratory) reference frame -- mostly connected with earth's surface. In these situations, CI has provided unique collapse hypersurface for a particular measurement. A relativistic generalization should posses this feature and at the same time, should be compatible with CI by definition and by the predicted results in all experiments performed so far. The proposed MLF collapse model, as the only one, satisfies these requirements.

The MLF model allows for construction of a measurement scenario, where two measurements on an uncollapsed polarization-entangled two-photon state is allowed. For this particular measurement, we consistently derive equations of the collapse lines, space-time characteristics of the experiment and point out the dependencies of all involved parameters. The adopted MLF model is shown to be compatible with CI of QS's collapse as far as all detectors are at rest in one reference frame, is independent of time sequence of measurements and, to our best knowledge, it is in agreement with all experimental observations so far. Similarly to H\&K model, the PLF and MLF model include a pre-collapse. In MLF model the pre-collapse occurs only in reference frames moving with respect to a detector taking a measurement and only on a part of the space-time. In consequence, MLF and H\&K models allows for taking multiple measurements on an uncollapsed QS, while the PLF (strictly speaking only in the preferred frame, see Appendix~A) and A\&A models do not. At the same time, the MLF and PLF models allows the initial quantum state to collapse in finite time before it gets into contact with any detector. This situation is briefly explored in~Appendix~A (for PLF model) and Appendix~E (for MLF model).

In this paper, the properties of the MLF collapse model are studied in detail with respect to space-like separated measurement on an polarization-entangled photon pair. Properties of the other collapse models in this measurement scheme are briefly examined in the Appendices~A~(PLF), C~(H\&K) and D(A\&A)~. In summary, there are four space-time models of QS's collapse examined in this paper:
 \begin{enumerate}
 \item Hellwig-Kraus (H\&K) -- along the backward light cone \cite{Hellwig1970formal}.
 \item preferred Lorentz frame (PLF) -- along space-like hyper surface $t=\mbox{const.}$ in unique preferred Lorentz frame \cite{Cohen1995retrodiction}.
 \item multiple Lorentz frame (MLF) -- along space-like hyper surface $t=\mbox{const.}$ in a rest frame of a detector \cite{Cohen1995retrodiction}.
 \item Aharonov-Albert (A\&A) -- along space-like hyper surface $t=\mbox{const.}$ in observer's Lorentz frame \cite{Aharonov1981can}/\footnote{Observer does not have to posses a detector and perform a measurement in this context.}.
 \end{enumerate}
 
The idea that in a system of two entangled particles, each particle can be measured before another has been proposed by Suarez~\cite{Suarez1997does}. He pointed out, that in this measurement scheme, measurements taken on both photons are uncorrelated. He used this statement as an axiom of his new measurement theory without proving its correctness or relation to the collapse models. In connection with this idea, he developed new measurement model called ``alternative description'' (AD). It was developed in order to deal with experiments involving impacts of photons on moving and stationary beam-splitters. Suarez used this description for prediction of measurement outcomes in experiment involving one moving and one stationary beam-splitter. Using AD, he predicted measurement outcomes which differ from prediction of quantum mechanics. The experiment has been modified and tested by Zbinden, Suarez~et.~al.~\cite{Zbinden2001experimental} with results agreeing with predictions of quantum mechanics.

Two space-like separated measurements on a two-particle quantum system in relation to time-ordering of the events has already been explored~\cite{Suarez1997does,Pater2021temporal}, but it has not been studied in relation to the PLF and MLF collapse models. We prove, that in the proposed experiment, the MLF model predict results distinct from A\&A and PLF models. It follows, the proposed experiment either validate or disprove the MLF model in favor of A\&A and PLF models. The possibility of multiple measurements on an uncollapsed QS in connection with the MLF model rises a few questions regarding the correlations of measured results and final states. Particularly, question of determination of the final state and question about violation of conservation laws in individual experiments. The questions are addressed in this paper.

There are two ways how to judge a compatibility of a collapse model with Copenhagen interpretation of a QS's collapse (CI). Either 1. By means of predicted results of a measurement or 2. By their definitions in a chosen Lorentz frame. 1. Until now, due to our best knowledge, all collapse models have revealed to predict the same results as CI, except the H\&K model~\cite{Cohen1995retrodiction}. This is considered in scenarios, where the detectors are stationary in the same reference frame. When a measuring detector (or an observer making prediction about a measurement) is in a movement, CI lacks explicit rules how to define the collapse hypersurface related to the measurement. In an experiment proposed in this paper, detectors in relative movement are considered to perform the measurements. It is demonstrated that MLF model differs in predicted results from A\&A and PLF models. Moreover, we show the H\&K model to predict result in contradiction with experiments~\cite{Salart2008,Scarani2000speed,Weihs1998violation} involving two space-like separated measurements on an entangled two-particle state. The measurements are taken by the stationary detectors. Thus, the experiments reveal H\&K model to be invalid. 

2.  A\&A and MLF models are compatible with CI by definition, while PLF and H\&K models are not. In practice, CI is used in a laboratory Lorentz frame, where all detectors taking measurements are at rest. Therefore, a collapse model is considered compatible if a collapse of a QS is instantaneous on all space in arbitrary laboratory Lorentz frame in which the detectors are at rest.
A\&A model is compatible with CI because the collapse hypersurfaces are instantaneous in any reference frame regardless of motion of the detectors. MLF model is compatible with CI because the measurements taken by the detectors, which are all stationary in arbitrary Lorentz frame, are related to instantaneous collapse hypersurfaces in the Lorentz frame. When one of the detectors in MLF model is in relative moment to the remaining detectors, compatibility of the model with CI breaks. We summarize, that both A\&A and MLF models have the same definition of collapse hypersurfaces, when all detectors are stationary in one reference frame. PLF model is compatible with CI only in the preferred Lorentz frame. Since the laboratory frame can in principle be associated with arbitrary Lorentz frame, the collapse hypersurfaces defined by the PLF model are not generally instantaneous in the laboratory frame. Thus, the collapse hypersurfaces in the PLF model are instantaneous only if the laboratory Lorentz frame is identical to the preferred Lorentz frame.  H\&K model does not agree with CI by definition in any reference frame, since the collapse of the QS precedes the measurement in any reference frame. Thus, PLF and H\&K collapse models are not compatible with CI, while MLF and A\&A are compatible with CI by definition.

In this article, we propose an experiment in which one moving and one stationary detector perform measurements on a quantum polarization-entangled photon-pair state. The measurements are assumed to be space-like separated. For analysis of this situation, we use MLF model, where the frame of instantaneous collapse is attached to a rest frame of a detector. We compare result of the experiment when MLF collapse model is used with results if H\&K, PLF and A\&A models are utilized. In Sec.~I an introduction into topic of collapse models was provided. In Sec.~II we introduce MLF collapse model along with collapse lines in detail and show its properties when a measurement is performed by one or two detectors. We focus on cases, when one of the detectors is in relative movement to other. The remaining collapse models are introduced and analyzed in Appendices~A~(PLF), C~(H\&K) and D~(A\&A). The experiment covering simultaneous measurement of two detectors on an uncollapsed state is analyzed in Sec.~III in relation to MLF model. The experiment is briefly discussed in connection to PLF, A\&A and H\&K models in Appendices~A, C and D, respectively. By careful inspection of the experiment through space-time diagrams and usage of the collapse lines, we identify origin of two temporary quantum states in the MLF model. In addition to this, we realize that a rule uniquely determining the final (permanent) state is missing in the quantum theory in combination with MLF model. From inspection of the space-time diagrams follows, that measurements on both photons are uncorrelated and can be interpreted as multiple measurements on the initial state. In Sec.~IV, the conservation laws are shown to be violated in the experiment, but not in mean value. In Appendix~B, we show how multiple measurements on uncollapsed state can be taken on a single photon state, which is in superposition of two distinct paths~\cite{Suarez2000preferred}. In Appendix~C, H\&K model is proven to provide result in contradiction with the experiments when two stationary detectors perform the measurements at two space-like separated events on an entangled state. The A\&A model and PLF model (in the preferred frame) are shown to hold the correlations between the photons. Thus, results of the proposed experiment either proves validity of MLF model or disproves it in favor of A\&A and PLF models. In Appendix~E, situation, when a pre-collapse occurs in MLF model is explored.

 \section{Relativity of a quantum state's collapse in MLF model}
 \label{sec2a}
 
 We inspect the space-time evolution of a QS $\vert \psi \rangle$ in one spatial $x$ and time dimension $t$. We work inside the Schr\"odinger picture and assume no interactions take place inside or outside the quantum system. Therefore, the QS changes only due to a collapse or an interaction-free evolution. The state is assumed to undergo a measurement at time $t = T_1$. Before the mea\-sure\-ment, we assume, that the quantum system is in the state $\vert \psi \rangle \equiv \vert \psi (t) \rangle$ on all available space $x$. After the measurement, the quantum system is in state $\vert \psi_{1,i} \rangle$, which is related to the original state $\vert \psi \rangle $ by projection postulate
 \begin{equation}
 \vert \psi_{1,i} \rangle = \dfrac{\langle \phi_i \vert \psi \rangle}{\vert \langle \phi_i \vert \psi \rangle \vert} \vert \phi_i \rangle.
 \label{e1}
 \end{equation}
The state $\vert \phi_i \rangle$ is an eigen-state of the measured observable's operator. If not explicitly stated otherwise, MLF collapse model is used throughout the paper.
\begin{figure}[!h]
\centering
\begin{subfigure}{0.45\textwidth}
\includegraphics[scale=1]{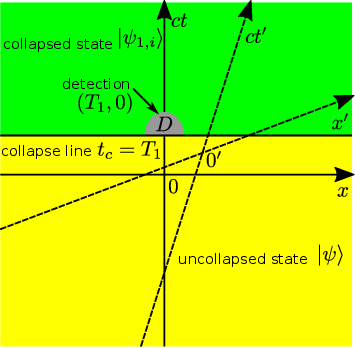}
\caption{}
\label{fig1a}
\end{subfigure}

\begin{subfigure}{0.45\textwidth}
\includegraphics[scale=1]{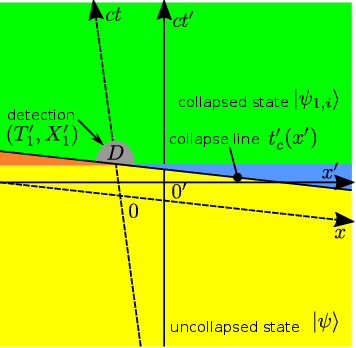}
\caption{}
\label{fig1b}
\end{subfigure}
\caption{(a) Space-time evolution of a QS $\vert
\psi \rangle$ utilizing the MLF collapse model. The QS $\vert
\psi \rangle$ undergoes a measurement at space-time point
$(t,x) = (T_1,0)$. The detector $D$, which performs the measurement, is in rest
in reference frame $S \equiv (ct,x)$. The
detector is placed at position $x = 0$. The yellow field denotes
the space-time area, where the QS $\vert \psi \rangle$ is
uncollapsed. The green area shows the space-time interval, where
the QS is collapsed to state $\vert \psi_{1,i} \rangle$. The $S'
\equiv (ct',x')$ reference frame moves relative to the reference
frame $S$ with positive velocity $v$. (b) Situation from
Fig.~\ref{fig1a} viewed from the reference frame $S'$. In reference frame $S'$ the QS $\vert \psi \rangle$ has undergone a measurement at space-time point $(T_1',X_1')$. The orange
area denotes space-time points in reference frame $S'$, where a
subsequent measurement by a detector, at rest in reference frame
$S'$, is allowed on an uncollapsed state $\vert \psi \rangle$. The blue region marks space-time points, where a the first measurement, by a detector in reference frame $S'$, is performed on already collapsed state.}
\label{fig1}
\end{figure}

 The space-time evolution of the QS $\vert \psi \rangle$, which in time $t = T_1$ undergoes a mea\-sure\-ment by a detector $D$ at position $x = 0$, is shown in Fig.~\ref{fig1a}. Evolution of the state  is shown in reference frame $S \equiv (ct,x)$. We assume, that the detector $D$ is in rest in the $S$ reference frame. In the Fig.~\ref{fig1a}, there are also depicted the axes $t'$ and $x'$ of a reference frame $S' \equiv (ct',x')$ moving relative to the reference frame $S$ with constant velocity $v>0$. The transformation of vector coordinates between reference frames $S$ and $S'$ are given by the Lorentz transformation
 \begin{eqnarray}
 \nonumber
 ct' &=& \gamma(ct - \beta x) + ct_0' \\
 x' &=& \gamma(x - \beta ct) + x_0'
 \label{e2}
 \end{eqnarray}
 and its inverse
 \begin{eqnarray}
 \nonumber
 ct &=& \gamma(ct' + \beta x') + ct_0 \\
 x &=& \gamma(x' + \beta ct') + x_0.
 \label{e3}
 \end{eqnarray}

 From Fig.~\ref{fig1a} follows, that the quantum system is on all space $x$ before the time of measurement $T_1$ in uncollapsed state $\vert \psi \rangle$/\footnote{According to Cohen and Hiley \cite{Cohen1995retrodiction}, assuming independent existence of a QS regardless of a measurement implies, that standard interpretation of quantum mechanics is being used.}. The space-time area, where the QS is uncollapsed has a yellow color. After a successful measurement by the detector $D$ at the space-time point $(T_1,0)$, the QS $\vert \psi \rangle$ collapses into the state $\vert \psi_{1,i} \rangle$ \cite{Susskind2014quantum} in the time of measurement $t = T_1$ on all space $x$, in accordance with Eq.~(\ref{e1}). The space-time area, where the QS is collapsed, is marked with the green color. It is important to note, that in practice there is not a fixed time in which the quantum system collapses. Rather, there is a time interval $\Delta T_1$ during which the quantum system interacts with the measurement apparatus and undergoes the collapse \cite{Garrisi2019experimental,Moreno2013looking}. For simplicity, we assume the collapse to occur at a fixed time $T_1$. As shown in Fig.~\ref{fig1a}, in reference frame $S$, the space-time line of the collapse (collapse line) is given by equation $t_c = T_1$.

 The same situation of the QS's collapse can be viewed from a reference frame $S' \equiv (ct',x')$, which moves relative to the detector $D$ (and reference frame $S$) with velocity $v > 0$. The space-time chart of this situation is shown in Fig.~\ref{fig1b}. It is apparent, that the collapse line $t_c'$ is not perpendicular to the time axis $t'$. It is tilted with non-zero angle with respect to the $x'$ axis according to the equation
 \begin{eqnarray}
\nonumber
 ct_c'(x') &=& - \beta x' + \dfrac{c T_1}{\gamma} + \beta x_0' + c t_0'\\ \nonumber
 &=& -\beta x' + \dfrac{cT_1'}{\gamma^2} + \beta x_0' + \beta^2 c t_0' \\
 &=& -\beta x' +c T_1' + \beta X_1'.
 \label{e4}
 \end{eqnarray}
 The measurement space-time point $(T_1,0)$ has been transformed into the $S'$ reference frame, to point $(T_1',X_1')$. The Eq.~(\ref{e4}) is a linear equation with respect to independent variable $x'$ with negative slope. The line related to the equation obviously passes through the point of measurement $(T_1',X_1')$, see Fig.~\ref{fig1b}. From the Eq.~(\ref{e4}) follows, that the ``speed'' of the QS's collapse triggered in reference frame $S$ is finite in reference frame $S'$ and equal to $-c/\beta$, which is in absolute value always grater than $c$.

  From an inspection of the Fig.~\ref{fig1b} follows, that after the measurement is taken in time $T_1'$, at point $X_1'$ by the detector $D$ at rest in reference frame $S$, there are places $x'$ at which a second measurement on the uncollapsed state $\vert \psi \rangle$ can be taken at a time $t'$ subsequent to the measurement time $T_1'$. The triangle-shaped area of the space-time points satisfying these requirements is depicted in Fig.~\ref{fig1b} by the orange color. The lower line of the triangle is given by the equation
  \begin{equation}
  t' = T_1',
  \end{equation}
  while the upper line of the triangle $t_c'(x')$ by Eq.~(\ref{e4}). Both lines have ending point in the measurement point $(T_1',X_1')$. In order to keep the measurement by detector $D$ unaffected, the second measurement has to be performed by a detector at rest in reference frame $S'$ at space-time point located in orange triangle in Fig.~\ref{fig1b}. In this scenario, both detectors take measurements on an uncollapsed QS $\vert \psi \rangle$. We develop this idea in detail at the end of this section. An evolution of the quantum state $\vert \psi \rangle$ subjected to a measurement by the detector $D$ with arbitrary constant velocity, as described in this paragraph, can be inspected as well by means of PLF, H\&K and A\&A collapse models. They are introduced in detail in Appendices~A,C and D, respectively. 
  
  If a second measurement in reference frame $S'$ was taken at space-time point in the blue area (see Fig.~\ref{fig1b}), the measurement is considered to be taken on the already collapsed state $\vert \psi \rangle$. The measurement, as observed in reference frame $S'$, is taken prior to the first measurement by detector $D$. For detailed discussion of this case, see Appendix~E. 

 The time interval $\Delta t_c'$ at position $x'$, in which a measurement on an uncollapsed state $\vert \psi \rangle$ can be taken in the moving reference frame $S'$, after the first measurement has been taken by the detector $D$ in the reference frame $S$ at space-time point $(T_1',X_1')$, is expressed by the equation
 \begin{eqnarray}
 \nonumber
 \Delta t_c'(x') &=& t_c'(x') - T_1' \\
 \nonumber
 &=& -\dfrac{\beta}{c}x' + \beta^2 (t_0' - T_1') + \dfrac{\beta x_0'}{c} \\
 &=& \dfrac{\beta}{c}(-x' + X_1').
 \label{e6}
 \end{eqnarray}
 From geometrical point of view, the time difference $\Delta t_c'$ expresses a height of the orange triangle in Fig.~\ref{fig1b} at arbitrary point $x'$. This expression is positive only at positions $x' < X_1'$, which is in agreement with placement of the orange triangle in Fig.~\ref{fig1b}. The spatial position of the detection point $X_1'$ (in reference frame $S'$) is equal to
 \begin{equation}
 X_1'  = -v(T_1' - t_0') + x_0'.
 \end{equation}

\begin{figure*}
 \centering
 \includegraphics[scale=1]{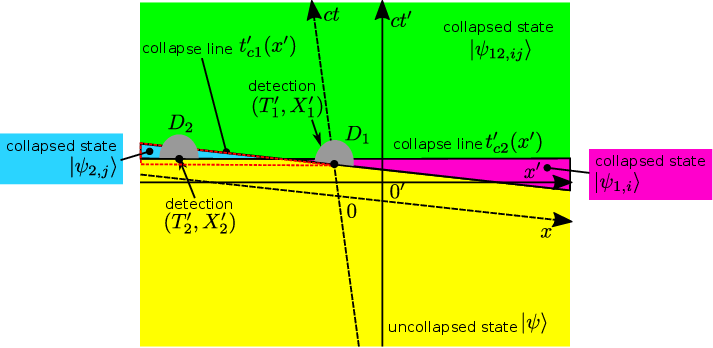}
 \caption{The space-time evolution of a QS $\vert \psi \rangle$, which collapses upon measurements by detectors $D_1$ and $D_2$, shown in reference frame $S' \equiv (ct',x')$. The reference frame $S \equiv (ct,x)$ moves in negative direction of axis $x'$. The collapse of QS $\vert \psi \rangle$ is triggered by two measurements. By detector $D_1$ at space-time point $(T_1',X_1')$ and detector $D_2$ at space-time point $(T_2',X_2')$. The measurement performed by detector $D_1$, at rest in reference frame $S$, divides the space-time diagram by collapse line $t_{c1}'$ to areas, where the QS is uncollapsed (yellow area) and collapsed to state $\vert \psi_{1,i} \rangle$ (magenta area). The measurement taken by detector $D_2$ divides the space-time diagram by collapse line $t_{c2}$ to areas, where the state is collapsed to state $\vert \psi_{2,j} \rangle$ (blue area) after the measurement and where the state is uncollapsed before the measurement. The triangle with red dashed border denotes area, where a second measurement can be performed on uncollapsed state $\vert \psi \rangle$ in reference frame $S'$ after the first measurement has been taken by detector $D_1$ at space-time point $(X_1',T_1')$ in reference frame $S$. The large green area in the upper part of the figure is given by intersection of collapse areas of detectors $D_1$ (magenta) and $D_2$ (blue). The QS, which occupy this area is $\vert \psi_{12,ij} \rangle$.}
 \label{fig2}
 \end{figure*}

In the Fig.~\ref{fig2} it is shown, how the second measurement performed by detector $D_2$ in rest in reference frame $S'$ at space-time point $(T_2',X_2')$ on the uncollapsed state $\vert \psi \rangle$ will affect its space-time evolution. Further, the detector $D$ will be denoted as $D_1$ and collapse line related to its measurement $t_c$ as $t_{c1}$. We assume, that detector $D_2$ remains at rest at fixed point $X_2'$ in reference frame $S'$. We further assume, that this detector performs a measurement, which causes the uncollapsed state $\vert \psi \rangle$ to be projected to state $\vert \psi_{2,j} \rangle$ upon detection of a value related to an eigen-vector $\vert \Psi_j \rangle$ according to projection postulate in Eq.~(\ref{e1}).  The detection performed by the detector $D_2$ is related to the collapse line $t_{c2}$. The line divides the
space-time to area, where the measured state $\vert \psi \rangle$ is uncollapsed (yellow area in Fig.~\ref{fig2}) and where it is collapsed after measurement by detector $D_2$ (union of blue and green areas).

There are four different collapse-related space-time regions which boundaries are given by collapse lines $t_{c1}$ and $t_{c2}$, see Fig.~\ref{fig2}. In the yellow space-time region the QS is in original uncollapsed state $\vert \psi \rangle$. The $\vert \psi_{1,i} \rangle$ state originates after detection by detector $D_1$. It occupies the magenta space-time region in the central-right part of the Fig.~\ref{fig2}. The state $\vert \psi_{1,i} \rangle$ is uniquely related to uncollapsed state $\vert \psi \rangle$ through the projection postulate in Eq.~(\ref{e1}). The same is valid for state $\vert \psi_{2,j} \rangle$ and the blue region in the central-left part of the Fig.~\ref{fig2}, but with relation to detector $D_2$. The QS $\vert \psi_{12,ij} \rangle$  emerges as the final collapsed state when both measurements (projections) by detectors $D_1$ and $D_2$ are taken into account. \textit{While the states $\vert \psi_{1,i} \rangle$ and $\vert \psi_{2,j} \rangle$ are uniquely given by the projection postulate, the final QS $\vert \psi_{12,ij} \rangle$ cannot be uniquely determined in some cases.} We will address this issue on an example of measurement of polarization-entangled photon-pair state in Sec.IV.

The detection points $(T_1',X_1')$ and $(T_2',X_2')$ in Fig.~\ref{fig2} are space-like separated. Therefore, the time sequence of detections by the detectors $D_1$ and $D_2$ is observer-dependent\footnote{ We can find one observer (moving with fixed pre-defined velocity), in which reference frame the time of detection of detector $D_1$ precedes detection by detector $D_2$. Example of this reference frame is $S'$ (see Fig.~\ref{fig2}). On the other hand, we can find another observer (moving with different fixed velocity), in which reference frame detection of detector $D_2$ precedes detection of detector $D_1$.  Example of this reference frame is $S$ (see Fig.~\ref{fig2}).}. It follows, that it is not possible to generally determine, which measurement triggered the collapse of the QS $\vert \psi \rangle$ first. But according to the space-time evolution diagram in Fig.~\ref{fig2}, it is possible to determine \textit{regardless of the reference frame}, which detector has been measuring on a collapsed or an uncollapsed QS.

 The scenario, when the measurements are taken by the two detectors $D_1$ and $D_2$ (as shown in Fig.~\ref{fig2} and Fig.~\ref{fig3}) has been investigated as well with utilization of PLF, H\&K and A\&A collapse models in Appendices~A,C and D, respectively. From the inspection with H\&K model follows, if both detectors $D_1$ and $D_2$ take space-like separated measurements, the measurements do not preserve correlations present in the initial state $\vert \psi \rangle$. This occurs regardless of the state of motion of both detectors. Since early 2000s, there have been experiments measuring correlations of entangled photon pairs at two space-like separated events with stationary detectors~\cite{Zbinden2001experimental,Salart2008,Weihs1998violation}. None of those experiments has reported violation of the correlations. Therefore, the H\&K model provides predictions conflicting with the experimental observations. On the other hand PLF and A\&A models predicts the correlations to be preserved regardless of the state of motion of the detectors, see Appendices A and D. These predictions is the opposite to MLF model in the proposed experiment. Therefore, the experiment involving multiple measurements on an uncollapsed state may either prove or disprove validity of the MLF collapse model.

 \section{Proposal of experiment involving two measurements on uncollapsed entangled photon-pair state}
\label{sec3}

We investigate the detection of a photon-pair state entangled in po\-la\-ri\-za\-tions and frequency according to the experimental layout in Fig.~\ref{fig2}. Conditions, by which measurements by detectors $D_1$ and $D_2$ are taken on an uncollapsed state are derived. We assume the detector $D_1$ to be located in a spatial origin $x = 0$ of a coordinate system $S \equiv(t,x)$ and the other $D_2$ in spatial origin $x' = 0$ of the coordinate system $S' \equiv (t',x')$. Detector $D_2$ moves relative to detector $D_1$ with positive velocity $v > 0$. The photon-pair source is assumed to be in rest in reference frame $S$. It will be resolved, at which experimental conditions two measurements on the uncollapsed QS will be allowed. 

\begin{figure}[!h]
 \centering
 \includegraphics[scale=1]{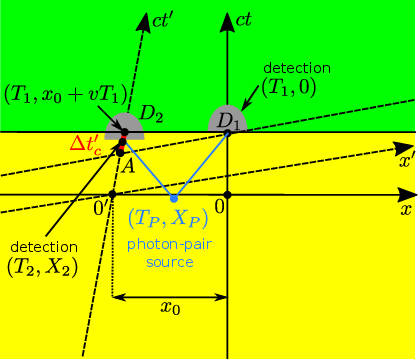}
 \caption{The space-time layout of the two-photon detection. A polarization-entangled photon pair is emitted by a source -- on the figure marked as blue dot. One photon propagates in positive direction of axis $x$ and the other in the negative direction of axis $x$ -- marked as blue lines. We assume the photon-pair source to remain in rest in reference frame $S \equiv (ct,x)$. Meanwhile, the detector $D_2$ moves in positive direction of axis $x$ with velocity $v$. The detector $D_2$ is located in spatial origin $x' = 0$ of reference frame $S' \equiv (ct',x')$. Detector $D_1$ remains at rest in spatial origin $x = 0$ of reference frame $S$. Detector $D_1$ detects a photon at space-time point $(T_1,0)$, while detector $D_2$ detects a photon in time interval $(T_1',T_1' + \Delta t_c')$, where $T_1'$ is time of detection $T_1$ expressed in reference frame $S'$ and $\Delta t_c'$ time difference given by Eq.~(\ref{e6}). By these conditions, both detectors measure on an uncollapsed two-photon state.}
 \label{fig3}
 \end{figure}

First, we define space-time scheme of the two-photon detection experiment, see Fig.~\ref{fig3}. Let us assume, that the detection of a photon by the detector $D_1$ occurs at time $T_1$. Further, we fix the constant $t_0$ in Eq.~(\ref{e3}) by requirement, that the origin $0' \equiv (t'=0,x'=0)$ of reference frame $S'$ has a time component in reference frame $S \equiv (t,x)$ equal to zero, $t=0$. In other words, we require the clocks at origins (detectors) in both frames $S$ and $S'$ to be synchronized at time $t' = t = 0$. By this assumption, using Eq.~(\ref{e21}) we have $t_0 = 0$. When the synchronization occurs, the origin $0'$ (of the reference frame $S'$) is located at position $x = x_0$ in reference frame $S$. The equation of motion of detector $D_2$ in reference frame $S$ is given by expression
\begin{equation}
x_{2}(t) = v t + x_0.
\label{e8}
\end{equation}

In reference frame $S$, the detector $D_2$ performs the measurement first at time $T_2$ in time interval
\begin{equation}
T_2 \in (T_1 - \Delta t_c, T_1).
\end{equation}
at position $X_2$, see Fig.~\ref{fig3}. In Fig.~\ref{fig3} the point of detection $(T_2,X_2)$ of detector $D_2$ on the uncollapsed state lies on the red line, which is a world-line of the detector $D_2$. The point of the earliest detection by detector $D_2$ is marked as $A$. The relation between the time differences $\Delta t_c'$, in Eq.~(\ref{e6}), and its counterpart $\Delta t_c$ in reference frame $S$ is given by equation for time dilatation
\begin{equation}
\Delta t_c = \gamma \Delta t_c'.
\label{e10}
\end{equation}
The time interval for time $T_2'$, when the detector $D_2$ performs a measurement on the uncollapsed state $\vert \psi \rangle$, can be expressed as
\begin{equation}
 T_2' \in (T_1', T_1' + \Delta t_c') \equiv (T_1',T_1' + \dfrac{\beta}{c} X_1'),
 \label{e11}
\end{equation}
which follows from Fig.~\ref{fig3}, Eq.~(\ref{e6}) and position of detector $D_2$ in reference frame $S'$ at coordinate $x'=0$. Since the right corner point of the interval has to be larger than the left one it follows, that $X_1' > 0$. $X_1'$ denotes position of detector $D_1$ in reference frame $S'$. This means, that in order to perform detection on the uncollapsed state, the detector $D_2$ has to be moving towards the detector $D_1$ from the negative side of axis $x$. Or in other words, the detector $D_1$ has to be ahead of detector $D_2$, positioned at positive coordinate $X_1'$, which is in agreement with the layout in Fig.~\ref{fig3}. The variables in the corner points of the interval in Eq.~(\ref{e11}) can be expressed by means of variables in reference frame $S$ as
\begin{eqnarray}
\label{e12}
T_1' &=& \gamma T_1 \\
\label{e13}
X_1' &=& \gamma(-x_0 - v T_1) = \gamma \Delta X_{12},
\end{eqnarray}
where $\Delta X_{12}$ is spatial distance between detectors $D_1$ and $D_2$ in time of detection $t=T_1$ of detector $D_1$ in reference frame $S$. From combination of Eqs.~(\ref{e6}), (\ref{e10}) and (\ref{e13}) follows, that the time difference $\Delta t_c$ is equal to
\begin{equation}
 \Delta t_c = \gamma^2\dfrac{\beta}{c}\Delta X_{12} \approx \dfrac{\beta}{c}\Delta X_{12} = \dfrac{v}{c^2} \Delta X_{12} \approx \Delta t_c'.
 \label{e14}
\end{equation}
The approximation in Eq.~(\ref{e14}) follows from assumption, that we neglect terms with powers in $\beta$ higher than one and keep only the first power of $\beta$ as the most significant contribution. The Eq.~(\ref{e14}) puts into relation the time differences $\Delta t_c'$ and $\Delta t_c$, during which a detection can be taken by detector $D_2$ on the uncollapsed state, velocity $v$ of detector $D_2$ and distance $\Delta X_{12}$ of detectors $D_1$ and $D_2$ in time $T_1$. Time $T_1$ is time of detection of a photon by detector $D_1$.

The QS $\vert \psi \rangle$ collapses after result of the measurement is stored in a classical memory of detector $D_2$. With regards to the current state of electronics, we estimate the time between detection and storage at the order of $10^{-10}\mbox{ s} = 0.1$~ns. Based on this assumption, the magnitude of product of the distance between detectors $\Delta X_{12}$ and velocity $v$ of detector $D_2$ can be derived from Eq.~(\ref{e14})
\begin{equation}
v \Delta X_{12} = c^2 \cdot 10^{-10} \mbox{s} \approx 10^{7} \mbox{m}^2\mbox{s}^{-1}.
\label{e15}
\end{equation}
If the moving detector was located in a vehicle moving on the earth's surface, the maximum allowed velocity $v$ could be at the order of $10^{2}$~m~s$^{-1}$ $= 360$~km~h$^{-1}$. When substituted into the Eq.~(\ref{e15}), the distance at which the photon pair should propagate $\Delta X_{12}$ is at the order $10^{5}~\mbox{m} = 100$~km. If the detector $D_2$ was placed on orbital station \cite{Pater2021temporal}, a two dimensional model in spatial coordinates considering non-collinear motion of the station and the photons would have to be developed\footnote{In this case, the time dilatation caused by general relativity effects would have to be considered as well.}.

The position $X_P$ and time of emission $T_P$ of the photon pair (see Fig.~\ref{fig3}) have to be properly tailored in order to hit the detectors $D_1$ and $D_2$ at required times $T_1$ and $T_2$/\footnote{In practice, the photon pair can be generated by a femto-second pump-beam pulse. It generates photon pair with pulse lengths in order of hundreds of femtoseconds \cite{Perina1999dispersion}. Therefore, the pulse lengths of photons in the pair are much smaller than the time required for detection and storage $\Delta t_s \sim 0.1$~ns. This guarantees successful detection of both photons in the pair at required times $T_1$ and $T_2$ with negligible error.}. We propose the time of detection $T_2$, by the moving detector $D_2$, to take place just after space-time point $A$ at coordinates 
\begin{eqnarray}
\label{e16}
T_2 &=& T_A + \varepsilon,\\
\label{e17}
X_2 &=& vT_2 + x_0 = X_A + v\varepsilon;\,\varepsilon > 0,
\end{eqnarray}
see Fig.~\ref{fig3}. The parameter $\varepsilon$ determines time difference between time of detection $T_2$ and time $T_A$. It transforms as time difference $\varepsilon = \gamma \varepsilon'$. This would provide detector $D_2$ the longest available time $\Delta t_c - \varepsilon$ for resolving of polarization of the photon and storage of the result. From this requirement and equations of motion for photons $x(t)=\pm c(t-T_P) + X_P$ follows, that the position $X_P$ and time $T_P$ have to obey the equations
\begin{eqnarray}
\label{e18}
X_P - c T_P &=& - c T_1,\\
\label{e19}
X_P + c T_P &=& X_2 + c T_2.
\end{eqnarray}
The Eqs.~(\ref{e18}) and (\ref{e19}) represent the set of two equations for two unknowns $X_P$ and $T_P$. Their solution is
\begin{eqnarray}
\label{e20}
X_P &=& \dfrac{X_2 + c T_2 - c T_1}{2}, \\[0.5cm]
\label{e21}
c T_P &=& \dfrac{X_2 + c T_2 + c T_1}{2}.
\end{eqnarray}
The set of Eqs.~(\ref{e20}) and (\ref{e21}) determine the position $X_P$ of the photon-pair source and time of their emission $T_P$ such that one photon from the pair hits detector $D_1$ at time $T_1$ at place $X_1$ and the other reaches moving detector $D_2$ at time $T_2$ at place $X_2$.

 The time $T_A$ and place $X_A$, which are related to time of detection $T_2$ and place of detection $X_2$ of detector $D_2$ through Eqs.~(\ref{e16}) and (\ref{e17}), are not in\-de\-pen\-dent parameters. Therefore we need to determine how they are related to the velocity $v$ of detector $D_2$, time of detection $T_1$ and position $x_0$ of detector $D_2$ at time $t = 0$. We have assumed, that the detector $D_2$ gets into contact with a photon in time $T_2 = T_A + \varepsilon$, where $T_A$ is time related to space-time point $A$, shown in Fig.~\ref{fig3}. From Fig.~\ref{fig3} follows, that $T_A' = T_1'$ at point $x = 0$. From this, we can derive space-time coordinates of point $(T_A,X_A)$ in reference frame $S$:
 \begin{eqnarray}
\label{e22}
cT_A &=& \gamma^2(cT_1 + \beta x_0), \\
\label{e23}
X_A &=& \beta \gamma^2(cT_1 + \beta x_0) + x_0 = \gamma^2(v T_1 + x_0).
\end{eqnarray}
The Eq.~(\ref{e23}) for point $X_A$ can be verified by utilization
of Eq.~(\ref{e6}) and Eq.~(\ref{e8}) of motion  for spatial
origin $x'= 0$ of reference frame $S'$ in reference frame $S$. By
setting $t = T_1 - \Delta t_c$ in the equation of motion
\begin{equation}
X_A = x_2(T_1 - \Delta t_c),
\end{equation}
the Eq.~(\ref{e23}) emerges as well.

From Eqs.~(\ref{e20}), (\ref{e21}), (\ref{e22}) and (\ref{e23}) follows, that the only independent parameters of the experiment are the velocity $v$ of the detector $D_2$, its position $x_0$ at time $t=0$ and time $T_1$ of detection of the photon by detector $D_1$. The space-time point $(T_2,X_2)$ of detection of detector $D_2$ is set by means of these parameters through Eqs.(\ref{e16}), (\ref{e17}), (\ref{e22}) and (\ref{e23}). The space-time point of emission of the photon pair $(T_P,X_P)$ is determined by the independent parameters as well in Eqs.~(\ref{e20}) and (\ref{e21}).

Experiment similar to the proposed one has been performed by Zbinden~et~al.~\cite{Zbinden2001experimental}, with regards to theoretical works of Suarez~\cite{Suarez1997does,Suarez1997relativistic}. Particularly, the goal of the experiment was to verify Suarez's theoretical model of measurement called AD. It predicted results distinct from quantum mechanics. The experiment disproved the AD theory in favor of well established quantum mechanics. The authors did not perform measurements, where both signal and idler photons' polarizations (or frequencies) were resolved. Instead, they left properties of the measured idler photon as unresolved. In Appendix~B, we have scatched an experiment proposed by Suarez~\cite{Suarez2000preferred}. In this experiment, two measurements on an uncollapsed \textit{single-photon state} $\vert \psi \rangle$ are performed. The analysis of this experiment can be made in the same way as the analysis of the experiment utilizing the photon-pair in Sec.III.

\section{Consequences of multiple measurements on uncollapsed two-photon state}
\label{sec4}
The QS $\vert \psi \rangle$ subjected to detection is maximally entangled state in polarizations $V$ and $H$ \cite{Arkhipov2015comparative,Perina2016spatial}
\begin{align}
\nonumber
\vert \psi \rangle = \iint_0^{\infty} \dfrac{d\omega_s d\omega_i}{\sqrt{2}}e^{i(\omega_s + \omega_i)t}\,\phi(\omega_s,\omega_i) & \left[  \vert V(\omega_s) \rangle_1 \vert H(\omega_i) \rangle_2 \right.\\
\label{e25}
+& \left. \vert H(\omega_s) \rangle_1 \vert V(\omega_i) \rangle_2 \right].
\end{align}
$\phi$ denotes two-photon amplitude. The $H$ and $V$ polarization states are assumed to be orthogonal to each other and related to transversally polarized photons. Let us assume, that photon state labeled by subscript $1$ is related to photon mode $1$ propagating in the positive direction of axis $x$. In the same way, the photon state labeled with subscript $2$ is related to photon mode $2$ propagating in negative direction of axis $x$. The polarization of the photon mode $1$ is resolved on detector $D_1$ and polarization of photon in mode $2$ by detector $D_2$. The maximally entangled QS in Eq.~(\ref{e25}) has been utilized in order to show the violation of quantum correlations, contained in the entanglement, in the measurement scheme proposed in Sec.~\ref{sec3}.

The polarizations measured by detectors $D_1$ and $D_2$ on state $\vert \psi \rangle$ in Eq.~(\ref{e25}) can be uncorrelated, if the two measurements are performed on the uncollapsed state. If the detectors $D_1$ and $D_2$ were placed in rest with respect to each other, the first polarization measurement by detector $D_1$, would project the state $\vert \psi \rangle$ into state, where only one predictable result can be obtained by polarization measurement with detector $D_2$. The polarizations obtained by the detectors would be orthogonal to each other. On the other hand, if we assume the detector $D_2$ measuring mode $2$ to be moving and the measurement scheme is obeyed as proposed, both detectors $D_1$ and $D_2$ are measuring on the uncollapsed state $\vert \psi \rangle$. This means, that both detectors have equal chance of measurement of $V$ polarized photon as well as horizontally polarized photon $H$. Therefore, there is 50~\% probability, that both photons would be measured with the same polarization. In this case, the measured photon-pair state would be $\vert V \rangle_1 \vert V \rangle_2$ or $\vert H \rangle_1 \vert H \rangle_2$. These measurement results are equivalent to situation when each detector performs a measurement on its own copy of an uncollapsed QS.

If polarization measurement (either by detector $D_1$ or $D_2$) is taken on one of the photons in state $\vert \psi \rangle$ in Eq.~(\ref{e25}), polarization of the other photon immediately after the measurement is uniquely given but not at arbitrary subsequent time. Let us assume a situation, when both detectors $D_1$ and $D_2$ measured $V$ polarization without frequency discrimination. For now, we assume, that both photons have been detected non-destructively \cite{Reiserer2013nondestructive,Niemietz2021nondestructive}. From inspection of the Fig.~\ref{fig2} follows, that the states $\vert \psi_{1,i} \rangle$ and $\vert \psi_{2,j} \rangle$ are determined with the projection postulate in Eq.~(\ref{e1}) from state $\vert \psi \rangle$ in Eq.~(\ref{e25}) utilizing projectors $\hat{P} = \int d\omega\,\vert V(\omega) \rangle_1 {}_1\langle V(\omega) \vert$ and $\hat{P} = \int d\omega\,\vert V(\omega) \rangle_2 {}_2 \langle V(\omega) \vert$, respectively. Since the frequency $\omega$ of the measured photon is not resolved, integration over variable $\omega$ is carried out. In this case, the projected states $\vert \psi_{1,V} \rangle$ and $\vert \psi_{2,V} \rangle$ (index of a state is related to mode and polarization of measured photon) are equal to
\begin{align}
\nonumber
\vert \psi_{1,V} \rangle &= \iint_0^{\infty} d\omega_s d\omega_i\,  \phi(\omega_s,\omega_i)e^{i(\omega_s+\omega_i)t}\\
\label{e26}
& \times \vert V(\omega_s) \rangle_1 \vert H(\omega_i) \rangle_2\\
\nonumber
\vert \psi_{2,V} \rangle &= \iint_0^{\infty} d\omega_s d\omega_i\,  \phi(\omega_s,\omega_i)e^{i(\omega_s+\omega_i)t}\\
\label{e27}
& \times \vert H(\omega_s) \rangle_1 \vert V(\omega_i) \rangle_2.
\end{align}
According to Fig.~\ref{fig2}, the states $\vert \psi_{1,V} \rangle$ and $\vert \psi_{2,V} \rangle$ are temporary. On the other hand, the final state $\vert \psi_{12,VV} \rangle$ (see Fig.~\ref{fig2}) is permanent from time point of view. \textit{But there is no rule by which this state can be determined.} The state $\vert \psi_{12,VV} \rangle$ cannot simply be defined as either $\vert \psi_{1,V} \rangle$ or $\vert \psi_{2,V} \rangle$, since they are different and there is no reason to prefer one against another. It can be suggested, that the final state $\vert \psi_{12,VV} \rangle$  should be given by results of the measurements -- the vertical polarizations of both photons by assumption, but this had to be verified by the experiment.

The possibility of obtaining the uncorrelated measurement results from state $\vert \psi \rangle$ in Eq.~(\ref{e25}) leads to violation of conservation laws, but not in mean value. We demonstrate this on energy conservation law. If the photon pair originated in process of spontaneous parametric down-conversion, the photons in modes 1 and 2 have the sum frequency equal to frequency of the pump beam
\begin{equation}
\hbar \omega_p = \hbar \omega_s + \hbar \omega_i.
\label{e28}
\end{equation}
At conditions, when both detectors $D_1$ and $D_2$ are at rest, measured frequencies of both photons $\omega_s$ and $\omega_i$ have to obey this equation. But in case, when both detectors $D_1$ and $D_2$ measure frequency on uncollapsed state $\vert \psi \rangle$, the energy conservation law in Eq.~(\ref{e28}) has to be obeyed independently by frequency $\omega_s$ and $\omega_i$. This follows from statement, that both measurements on the QS $\vert \psi \rangle$ are uncorrelated. In consequence, \textit{the sum of measured frequencies $\omega_s$ and $\omega_i$ does not have to be equal to $\omega_p$ in individual measurements.}

The probability density $p(\omega_s,\omega_i)$ of detecting signal photon with frequency $\omega_s$ and idler photon with frequency $\omega_i$, when both detectors $D_1$ and $D_2$ are at rest, is equal to
\begin{equation}
p(\omega_s,\omega_i) = \vert \phi(\omega_s,\omega_i) \vert^2.
\label{e29}
\end{equation}
Contrary, if frequencies of both photons $\omega_s$ and $\omega_i$ are measured on the uncollapsed state, the probability density $p_N(\omega_s,\omega_i)$  of detection of signal photon with frequency $\omega_s$ and idler photon with frequency $\omega_i$ on uncollapsed state is given by equation
\begin{equation}
p_N(\omega_s,\omega_i) =  \int_0^\infty d\bar{\omega}_i\, p(\omega_s,\bar{\omega}_i) \int_0^\infty d\bar{\omega}_s\, p(\bar{\omega}_s,\omega_i).
\label{e30}
\end{equation}
The different expression for probability density $p_N$ emerges from an assumption, that when frequency of one of the photons is measured on an uncollapsed state, frequency of the other is still not determined. The probability density $p_N(\omega_s,\omega_i)$ is function separable in variables $\omega_s$ and $\omega_i$, while function $p(\omega_s,\omega_i)$ is non-separable.

The mean value of energy $\langle E \rangle$ in QS $\vert \psi \rangle$ measured by static detectors is equal to
\begin{equation}
\langle E \rangle = \hbar\langle \omega_s \rangle_p + \hbar \langle \omega_i \rangle_p.
\end{equation}
The subscript $p$ in the mean values $\langle \omega_s \rangle_p$ and $\langle \omega_i \rangle_p$ denotes averaging with respect to probability density $p$ defined in Eq.~(\ref{e29}). It follows, that the mean value of energy $\langle E \rangle$ in the $\vert \psi \rangle$ state divided by $\hbar$ is given by sum of the mean values $\langle \omega_s \rangle_p$ and $\langle \omega_i \rangle_p$ of measured frequencies $\omega_s$ and $\omega_i$ computed with probability density $p = p(\omega_s,\omega_i)$ defined in Eq.~(\ref{e29}). If the frequencies $\omega_s$ and $\omega_i$ are both measured on the uncollapsed state $\vert \psi \rangle$ the mean value of energy $\langle E \rangle$ in QS $\vert \psi \rangle$ has to be computed with probability density $p_N(\omega_s,\omega_i)$ defined in Eq.~(\ref{e30}). It is straightforward to show, that mean value of measured signal frequency $\omega_s$ is the same for both probability densities $p_N$ and $p$/\footnote{To prove this statement, normalization condition for the probability density $p$, $\iint_0^\infty d \omega_s d \omega_i\, p (\omega_s,\omega_i) = 1$ has to be used.}. This holds for the mean value of the idler frequency $\omega_i$ as well. Therefore the mean value of energy $\langle E \rangle$ in state $\vert \psi \rangle$ is the same regardless of utilization of multiple measurements on the uncollapsed state.

\section{Conclusion}
We have shown the diagrams of space-time evolution of a general quantum state subjected to a measurement by a detector in the reference frame of the detector and in the reference frame moving with constant velocity relative to the detector. The evolution has been studied with utilization of Multiple Lorentz frame (MLF), Preferred Lorentz frame (PLF), Hellwig-Kraus (H\&K) and Aharonov-Albert (A\&A) collapse model. The evolution has been analyzed as well in cases when the measurement is taken by two detectors, which detections are space-like separated. Compatibility of all models with Copenhagen Interpretation has been examined. From the MLF space-time model of quantum state's collapse follows, that multiple measurements can be taken on the same uncollapsed quantum state. In the opposite, PLF and A\&A models revealed to lack this type of measurement in the considered measurement scheme. Hellwig-Kraus model has been proved to provide invalid result in measurement scenarios of this type. Particularly, it predicts results conflicting with experiments involving space-like separated measurements

With MLF model we show that under certain conditions one stationary and one moving detector can perform measurements on uncollapsed photon-pair state entangled in polarizations and energy. Allowance of multiple measurements on the uncollapsed quantum state results in uncorrelated measurement results even if the uncollapsed state is maximally correlated. This is in opposite with PLF and A\&A model, where the measurements are correlated. With PLF collapse model, all predictions have to be made in the preferred frame in order to avoid back-in-time collapses affecting results of the former measurements. The uncorrelated measurements in MLF model lead to violation of conservation laws and unavailability to determine the collapsed quantum state, which originates after all detections take place. On example of frequency measurement on the photon-pair state, it is discussed, that the energy conservation is violated in individual measurements but not in mean value. The result of the proposed two-photon experiment determines wether the MLF model is valid while leaving the A\&A and PLF models generally invalid. An alternative less demanding experiment which allows for multiple measurement on a single photon state has been scatched. It involves single-photon state superposed in two distinct paths. 

The proposed photon-pair experiment can be generalized on entangled multi-photon state with multiple detectors, each moving with its own velocity. As a consequence, multiple measurements -- more than two, on a uncollapsed quantum state are allowed in the presented MLF collapse model.

\section{Acknowledgements}
The author gratefully acknowledges J.~Peřina~Jr. and L.~Richterek from Palacký~University, Faculty~of~Sciences for useful dis\-cu\-ssions, corrections and editing of the manuscript.

\appendix
\section{Preferred Lorentz frame collapse model}
\begin{figure}[!h]
\begin{subfigure}{0.45\textwidth}
\includegraphics[scale=1]{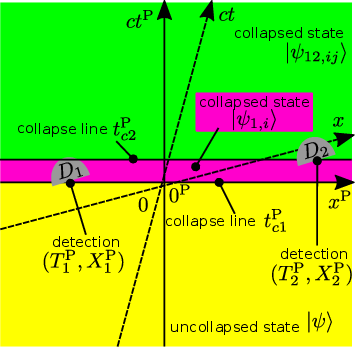}
\caption{}
\label{fig4a}
\end{subfigure}
\begin{subfigure}{0.45\textwidth}
\includegraphics[scale=1]{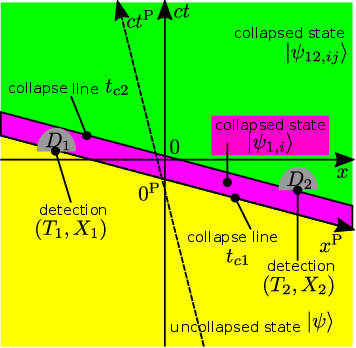}
\caption{}
\label{fig4b}
\end{subfigure}
\caption{(a) Space-time evolution of a QS $\vert \psi \rangle$ viewed from preferred Lorentz frame $S^{\rm P} \equiv (ct^{\rm P},x^{\rm P})$. For determination of the collapse lines $t_{c1}^{\rm P}$ and $t_{c2}^{\rm P}$ the PLF collapse model is utilized. The initial QS $\vert \psi \rangle$ undergoes two measurements at space-time points $(T_1^{\rm P}, X_1^{\rm P})$ and $(T_2^{\rm P}, X_2^{\rm P})$ by detectors $D_1$ and $D_2$, respectively. Both detectors are assumed to be at rest in reference frame $S \equiv (ct,x)$, which axes are as well depicted. The reference frame $S$ propagates along positive direction of axis $x^{\rm P}$ with constant velocity $v$. After measurement by detector $D_1$, the QS $\vert \psi \rangle$ collapses to QS $\vert \psi_{1,i} \rangle$, which occupies the magenta space-time region. After measurement by detector $D_2$, the QS $\vert \psi_{1,i} \rangle$ collapses to final QS $\vert \psi_{12,ij} \rangle$, which is located in the green space-time region. The magenta box in which the caption of the QS $\vert \psi_{1,i} \rangle$ is placed only highlights affiliation of the caption to the magenta space-time region. (b) Space-time evolution of a QS $\vert \psi \rangle$  in Fig.~\ref{fig4a}, viewed from reference frame $S \equiv (ct,x)$. Collapse line $t_{c1}$ ($t_{c2}$) is collapse line $t_{c1}^{\rm P}$ ($t_{c2}^{\rm P}$) viewed from reference frame $S$.}
\label{fig4}
\end{figure}

In preferred Lorentz frame (PLF) collapse model, collapse of a QS is instantaneous in one unique Lorentz frame. It will be denoted as $S^{\rm P} \equiv (ct^{\rm P},x^{\rm P})$. For an example, please see Fig.~\ref{fig4a}, where two collapses, triggered by the measurements of detectors $D_1$ and $D_2$, occur. This means, that in all other Lorentz frames, a collapse hypersurface will be a space-like hypersurface spanning both into the past and future~\cite{Cohen1995retrodiction}. For an example, see Fig.~\ref{fig4b}, where the collapse lines are viewed from reference frame $S \equiv (ct,x)$ moving with positive constant velocity along axis $x^{\rm P}$. The preferred Lorentz frame can be associated with center of mass of entangled two-particle system or microwave background radiation~\cite{Cohen1995retrodiction}. We remark, that in the preferred Lorentz frame $S^{\rm P}$, orientation of the collapse hypersurface is not dependent on speed of an observer either taking measurement with a detector or making a prediction of a measurement. In a result, for the observer at rest in preferred Lorentz frame $S^{\rm P}$, PLF and A\&A collapse models are equivalent. As far as all the detectors are at rest in preferred Lorentz frame and observers are equipped with detectors, all the studied models except H\&K (PLF, MLF and A\&A) are equivalent with respect to predicted results of measurements, orientation of collapse hypersurfaces (lines) and compatibility with CI. 

Let us examine an experimental scenario, when two detectors $D_1$ and $D_2$ are taking space-like separated measurements on the polarization-entangled photon-pair state in Eq.~(\ref{e25}). Since PLF collapse hypersurfaces' orientation is the same regardless of motion of the detectors, for simplicity we will assume both of the detectors ($D_1$ and $D_2$) to be at rest in reference frame $S \equiv (ct,x)$. We further assume the reference frame $S$ to propagate in positive direction of axis $x^{\rm P}$ with constant velocity $v$. The examined experimental situation viewed from reference frame $S$ and $S^{\rm P}$ is shown in Figs.~\ref{fig4a} and~\ref{fig4b}. We note, that the detection events $(T_1^{\rm P},X_1^{\rm P})$ and $(T_2^{\rm P},X_2^{\rm P}); T_1^{\rm P}~<~T_2^{\rm P}$ are placed in such a way, that they change their temporal order, when viewed from reference frame $S;\,T_1 > T_2$ (see Figs.~\ref{fig4a} and \ref{fig4b}). 

The measurement scheme described from the Lorentz preferred frame $S^{\rm P} \equiv (ct^{\rm P},x^{\rm P})$ (see Fig.~\ref{fig4a}) resembles a familiar scheme obtained with Copenhagen Interpretation. But in this situation, both detectors $D_1$ and $D_2$ are in a movement. In the scheme the initial QS $\vert \psi \rangle$ undergoes two measurements. First at point $(T_1^{\rm P},X_1^{\rm P})$ by moving detector $D_1$, where the QS collapses from initial state $\vert \psi \rangle$ (defined at all points of yellow region) to state $\vert \psi_{1,i} \rangle$ (defined at each point of magenta region). Then, a second measurement by detector $D_2$ at space-time point $(T_2^{\rm P},X_2^{\rm P})$ is performed with associated collapse of state $\vert \psi_{1,i} \rangle$ to final state $\vert \psi_{12,ij} \rangle$. From this perspective, all studied phenomena are in full agreement with CI.

When the same situation as in Fig.~\ref{fig4a} is inspected from reference frame $S$, the temporal order of measurement at spatial points $X_1$ and $X_2$ reverses ($T_1 > T_2$). In such case, we may experience a collapse of a subsequent measurement at space-time point $(T_1,X_1)$ affecting previously measured result at space-time point $(T_2,X_2)$. What is depicted in Fig.~\ref{fig4b} is the final situation, when both measurements by detectors $D_1$ and $D_2$ are taken into account.  Let us analyze the situation depicted in Fig.~\ref{fig4b} from start step by step. First, we assume, that there is an initial state $\vert \psi \rangle$ defined by Eq.~(\ref{e25}) undergoing a measurement with detector $D_2$ at space-time point $(T_2,X_2)$. This triggers collapse associated with collapse line $t_{c2}$. This collapse line is instantaneous in preferred Lorentz frame $S^{\rm P}$. We stress that the collapse line $t_{c2}$ leads above the second detection point $(T_1,X_1)$ (see Fig.~\ref{fig4b}) and leaves the QS at this point unaffected by the measurement. After measurement by the detector $D_2$, the state $\vert \psi \rangle$ collapses to state $\vert \psi_{2,j} \rangle = \langle \psi \vert \Psi_{2,j} \rangle \vert \Psi_{2,j} \rangle$, where the state $\vert \Psi_{2,j} \rangle$ is associated with the value measured by the detector $D_2$. For simplicity, the state $\vert \psi_{2,j} \rangle$ is kept un-normalized. State $\vert \psi_{2,j} \rangle$ will temporarily occupy the green region in Fig.~\ref{fig4b}, until a second measurement after time $T_1 - T_2$ at space-time point $(T_1,X_1)$ is performed. 

At time $T_1$, subsequent to time $T_2$ in reference frame $S$ (see Fig.~\ref{fig4b}), a measurement by detector $D_1$ at space-time point $(T_1,X_1)$ is performed. This measurement takes place on the uncollapsed state $\vert \psi \rangle$, as explained in the previous paragraph. Therefore, it is tempting to arrive to a conclusion that the measurements by the detectors $D_1$ and $D_2$ can be uncorrelated; Since both of them are taken on the initial entangled state $\vert \psi \rangle$. But what has to be taken into account is the backward-in-time propagating collapse line $t_{c1}$ associated with the letter measurement by detector $D_1$ at space-time point $(T_1,X_1)$. This collapse line causes the initial QS $\vert \psi \rangle$ to collapse at point of first measurement $(T_2,X_2)$ and even in finite time before the first measurement takes place to state $\vert \psi_{1,i} \rangle$ (it occupies the magenta region in Fig.~\ref{fig4b}). This collapse (associated with collapse line $t_{c1}$) ensures, that the measurements at both space-time points $(T_2,X_2)$ and $(T_1,X_1)$ are correlated, although the measurement at space-time point $(T_2,X_2)$ has been initially assumed to be taken on the uncollapsed state $\vert \psi \rangle$. In consequence, in the green region of Fig.~\ref{fig4b}, the final state $\vert \psi_{12,ij} \rangle$ emerges. It takes both projective measurements by detectors $D_1$ and $D_2$ into account. 

This backward-in-time collapse (associated with line $t_{c1}$) in reference frame $S$ may affect a polarization value initially measured by detector $D_2$. This occurs, because the measurement scenario in Fig.~\ref{fig4a} is inspected in reference frame $S$, where the measurement points $(T_1,X_1)$ and $(T_2,X_2)$ exchange their temporal order (see Figs.~\ref{fig4a}~and~\ref{fig4b}). It is hard to accept that a measurement, which occurs first (by detector $D_2$), may be affected by a measurement after it (by detector $D_1$). Particularly, it can be assumed that detector $D_2$ measured $V$ polarization (with no frequency discrimination) first (see Fig.~\ref{fig4b}). Then, at time $T_1 > T_2$, a subsequent measurement is performed by detector $D_2$ on the initial state $\vert \psi \rangle$ as well. This measurement may result into polarization $V$ as well. But the letter measurement causes collapse backward in time $t$, which changes the polarization measured by detector $D_1$ from $V$ to $H$, in order to keep correlations established in the initial state $\vert \psi \rangle$. Thus, all predictions of measurement outcomes with PLF model have to be made in the preferred frame $S^{\rm P}$ in order to avoid backward-in-time collapse situations, which are hard to interpret. In the preferred Lorentz frame $S^{\rm P}$, the measurements by detectors $D_1$ and $D_2$ are always correlated. Therefore, the predicted results of measurements by PLF and A\&A~(see Appendix~D) collapse models on the polarization entangled state $\vert \psi \rangle$ in Eq.~(\ref{e25}) are the same. Backward-in-time collapse lines appear in the MLF model and H\&K model as well. But they do not enforce correlations backward in time like in the case of PLF model. The H\&K and MLF models either leaves both measured results uncorrelated or keep the measured results correlated since the initial measurement, in dependence on the particular experimental scheme.

\section{Multiple measurements on an uncollapsed single photon state superposed in two distinct paths}
In Sec.~\ref{sec3} experimental scenario involving multiple measurements on an uncollapsed photon-pair state entangled in energy and polarization has been described. This type of experiment requires generation of a photon-pair state and utilization of both photons during the non-collapsing measurements. The generation of a photon-pair state is a second-order process. Thus, the generation rate of the photon-pairs is low in comparison with the first-order processes. Moreover, generation of a photon-pair simultaneously entangled in energy and polarization, as required by the proposed experiment (see Eq.~(\ref{e25})), can be challenging. Therefore, we briefly outline an experiment in which multiple measurements on an uncollapsed state is performed on a single photon state superposed in two distinct paths. Since the experiment does not include entangled particles, it does not show consequences of the uncollapsing measurements on the correlations. It is shown, that the uncollapsing measurement on the single photon state superposed in two distinct paths may result in cloning of the single photon state or its complete destruction. This experiment has been already proposed by Suarez~\cite{Suarez2000preferred}.

We assume, that the initial single photon state $\vert \psi \rangle$ is superposed both in frequency $\omega$ and paths l1 and l2 according to formula
\begin{equation}
\vert \psi \rangle = \int_0^\infty d\omega\, \dfrac{\phi(\omega)}{\sqrt{2}} a_{\rm l1}^\dagger(\omega) e^{i\omega t} \vert 0  \rangle + \dfrac{\phi(\omega)}{\sqrt{2}} a_{\rm l2}^\dagger(\omega) e^{i \omega t} \vert 0 \rangle.
\label{eA1}
\end{equation}
The creation operator $a^\dagger_{\rm l1}(\omega)$ ($a^\dagger_{\rm l1}(\omega)$) creates a photon with frequency $\omega$ in path l1 (l2). The state is superposed in two distinct paths denoted as l1 and l2, since the multiple measurements on an uncollapsed  photon state requires the photon to be on two distinct places in the same time. The QS in Eq.~(\ref{eA1}) can be obtained by sending a single photon QS $\vert \Phi \rangle = \int_0^\infty d\omega\, \phi(\omega) a^\dagger(\omega) e^{i\omega t} \vert 0 \rangle$ to a 50:50 beam-splitter with l1 and l2 paths on its outputs. The superposition in frequency $\omega$ in state $\vert \psi \rangle$ in Eq.~(\ref{eA1}) weighted by a function $\phi(\omega)$ emerges from requirement of the photon to be localized in time pulse narrower than $10^{-10}$~s.

The space-time diagram of propagating of the photon along the paths l1 and l2 including measurement is analogical to Fig.~\ref{fig3}. Path l1 can be connected with path of the photon propagating in negative direction of axis $x$ and path l2 with path of a the photon propagating in positive direction of axis $x$. The scheme of measurement by detectors $D_1$ (on path l1) and $D_2$ (on path l2) remains the same. In analogy to photon-pair's case, the photon is observed by both detectors $D_1$ and $D_2$ with probability 1/4. If the detections by the detectors $D_1$ and $D_2$ were nondestructive, two real photons would propagate away from them after the detection. With the same probability no photon can be detected as well. This situation would result in destruction of both photons. In rest of the cases, a photon is detected either by detector $D_1$ or $D_2$, as usual. When two photons are observed simultaneously, they are clones with respect to all degrees of freedom except the spatial ones. 

\section{Hellwig-Kraus collapse model}
\label{sec2b}
\begin{figure}[!h]
\begin{subfigure}{0.45\textwidth}
\includegraphics[scale=1]{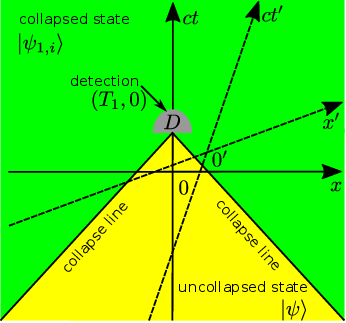}
\caption{}
\label{fig5a}
\end{subfigure}
\begin{subfigure}{0.45\textwidth}
\includegraphics[scale=1]{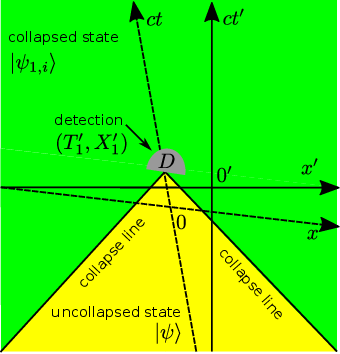}
\caption{}
\label{fig5b}
\end{subfigure}
\caption{(a) Space-time evolution of a QS $\vert \psi \rangle$ in reference frame $S\equiv (ct,x)$. The QS undergoes a measurement by a stationary detector $D$ at space-time point $(T_1,0)$. After the measurement the QS $\vert \psi \rangle$ collapses into state $\vert \psi_{1,i} \rangle$. The space-time distribution of the collapse is determined by H\&K model. The collapse takes place along the past light cone with apex in the measurement point $(T_1,0)$. The space-time regions of collapsed and uncollapsed states are divided by the collapse lines $x(t) = \pm c(t - T_1)$. Axes $x'$ and $ct'$ of reference frame $S' \equiv (ct',x')$ moving with positive velocity $v$ in reference frame $S$ are as well depicted. (b) The situation (a), but viewed from the reference frame $S'$.}
\label{fig5}
\end{figure}

Let us study a space-time evolution of a quantum state $\vert \psi \rangle$, which is measured at space-time point $(T_1,0)$ in reference frame $S$ (see Fig.~\ref{fig5a}).  The measurement triggers QS's collapse according to H\&K collapse model. It is assumed, that detector $D$ is at rest in reference frame $S$ at spatial coordinate $x=0$. H\&K
model requires the collapse lines to be aligned along the past light cone with apex in a measurement point, see Fig.~\ref{fig5a}.
Since the speed of light $c$ is invariant in all reference frames, a collapse line preserves its slope in all reference frames. For completeness, the process of measurement on the QS $\vert \psi \rangle$ viewed from moving reference frame $S'$ is shown in Fig.~\ref{fig5b}.

 H\&K model demands QS $\vert \psi \rangle$ to start its collapse before the measurement takes place, see Fig.~\ref{fig5a}. We admit, that the MLF and PLF model suffers of this problem as well. Utilizing MLF model (see Fig.~\ref{fig1b}), in reference frame moving relative to the detector $D$ a pre-collapse can be observed. Particularly, in space-time area $(x' > X_1) \wedge (t' < T_1')$. For detailed discussion of this case, please see Appendix~E. When considering the H\&K model again, the space-time point of measurement in the future is predetermined and cannot be avoided in any reference frame. Its sudden change would demand changing of the collapsed QS $\vert \psi_{1,i} \rangle$ to uncollapsed QS $\vert \psi \rangle$ in the past (see Fig.~\ref{fig1b}).

\begin{figure}[!h]
\includegraphics[scale=1]{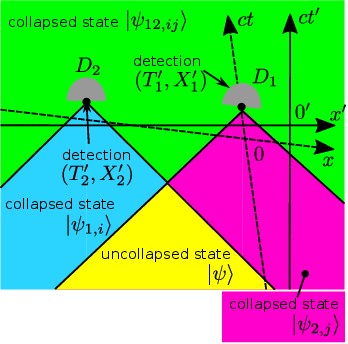}
\caption{Space-time evolution of a QS $\vert \psi \rangle$ subjected to two space-like separated measurements by detectors $D_1$ and $D_2$. In this diagram H\&K collapse model is used. The evolution is viewed in reference frame $S'$. The detections are taken at space-time points $(T_1',X_1')$ -- by $D_1$ and $(T_2',X_2')$ -- by $D_2$. Detector $D_1$ ($D_2$) remains at rest in reference frame $S\equiv(ct,x)$ ($S' \equiv(ct',x')$). Measurement by detector $D_1$ on state $\vert \psi \rangle$, causes its collapse to state $\vert \psi_{1,i} \rangle$. Accordingly, measurement of detector $D_2$ on state $\vert \psi \rangle$, causes its collapse to state $\vert \psi_{2,j} \rangle$. State $\vert \psi_{12,ij} \rangle$ originates by taking into account measurements of both detectors.}
\label{fig6}
\end{figure}

In Fig.~\ref{fig6} we investigate a situation, when two space-like separated detectors $D_1$ and $D_2$ perform the measurements on an initial state $\vert \psi \rangle$. We prove, that from definition of H\&K model immediately follows, that the measurements performed on two entangled particles \textit{are not} correlated, regardless of state of motion of the detectors~\footnote{In~\cite{Hellwig1970formal}, see Fig.~3 together with its description in paragraph above Eq.~(3) and definition in Eq.~(2).}. 
Let us keep the same assumptions as in analysis of the MLF model -- the detector $D_1$ ($D_2$) remains at rest in reference frame $S$ ($S'$). The reference frame $S'$ propagates along positive direction of axis $x$ with velocity $v>0$ in reference frame $S$. The space-time diagram of this situation in Fig.~\ref{fig6} is viewed from reference frame $S'$.

Initially, the state is prepared in state $\vert \psi \rangle$, then the measurements are taken by the detectors $D_1$ and $D_2$ (see Fig.~\ref{fig6}). Either detectors cause the QS to collapse outside their backward light cones. Area inside intersection of both backward light cones has yellow color and all its points are occupied by the initial QS $\vert \psi \rangle$. The area outside the backward light cone of detector $D_1$ intersected with backward light cone of detector $D_2$ has a blue color. This area is taken by QS $\vert \psi_{1,i} \rangle$. The QS $\vert \psi_{1,i} \rangle$ is created by measurement of detector $D_1$ on initial QS $\vert \psi \rangle$, due to H\&K model. Therefore, with regard to Eq.~(\ref{e1}), the relation between state $\vert \psi_{1,i} \rangle$ and $\vert \psi \rangle$ is given by the projection postulate
\begin{equation}
\vert \psi_{1,i} \rangle = {}_1\langle \phi_i \vert \psi \rangle \vert \phi_i \rangle_1.
\label{eB1}
\end{equation}
In Eq.~(\ref{eB1}), we have assumed that detector $D_1$ measured value $d_i$ related to eigen-state $\vert \phi_i \rangle$. For simplicity, we kept the state $\vert \psi_{1,i} \rangle$ in Eq.~(\ref{eB1}) unnormalized. In the same way, the magenta region related to state $\vert \psi_{2,j} \rangle$ originating by measurement of detector $D_2$ on initial QS $\vert \psi \rangle$ is defined by equation
\begin{equation}
\vert \psi_{2,j} \rangle = {}_2\langle \Psi_j \vert \psi \rangle \vert \Psi_j \rangle_2.
\label{eB2}
\end{equation}
The final state $\vert \psi_{12,ij} \rangle$ takes into account both measurements by detectors $D_1$ and $D_2$ and is equal to
\begin{equation}
\vert \psi_{12,ij} \rangle = {}_1\langle \phi_{i} \vert  {}_2\langle \Psi_{j} \vert \psi \rangle \vert \phi_{i} \rangle_1 \vert \Psi_{j} \rangle_2. 
\end{equation}
From Eqs.~(\ref{eB1})~and~(\ref{eB2}) immediately follows, that both measurements of detectors $D_1$ and $D_2$ take place on the initial uncollapsed state $\vert \psi \rangle$. Therefore, the measurements contradict with predictions of QM. Particularly, the measurements should be uncorrelated. Cohen and Hiley~\cite{Cohen1995retrodiction} arrived to the same conclusion. In addition, they suggested, that the H\&K model can be redefined such that both measurement by the detectors $D_1$ and $D_2$ are taken on already collapsed state $\vert \psi_{12,ij} \rangle$.

Since a backward light cone preserves its shape regardless of value of constant velocity of a detector, the uncorrelated measurements must be observed in experiments, where both detectors $D_1$ and $D_2$ are stationary and their detections are space-like separated. Fortunately, experiments of this type have already been performed~\cite{Salart2008,Scarani2000speed,Weihs1998violation}. Goal of all the cited works has been to prove violation of Bell's inequalities with space-like separated detections. Particularly, Salart~et.~al.~\cite{Salart2008} and Scarini~et.al.~\cite{Scarani2000speed} have been utilizing entanglement of photon pairs in time domain, while Weihs~et.al. have been utilizing entanglement in polarizations, as proposed in this paper. No contradictions with predictions of quantum-mechanics or violation of correlations have been reported. Therefore, H\&K collapse model should be considered as generally invalid.

\section{Aharonov-Albert collapse model}

A\&A collapse model requires a collapse of a QS to be instantaneous for any observer~\cite{Cohen1995retrodiction,Aharonov1981can}. In this model the observer does not have to be equipped with a detector. To introduce the model let us assume, as usual, a situation, where a single detector $D$ takes a measurement on QS $\vert \psi \rangle$ (see Fig.~\ref{fig1a}). The detector is assumed to be stationary in reference frame $S$ and the measurement occurs at space-time point $(T_1,0)$. Let us further assume, that the observer is associated with the detector $D$. Then, the space-time scheme of the measurement is the same like in Fig.~\ref{fig1a}. I.e. the collapse line related to measurement of detector $D$ is instantaneous in reference frame $S$.

\begin{figure}
\includegraphics[scale=1]{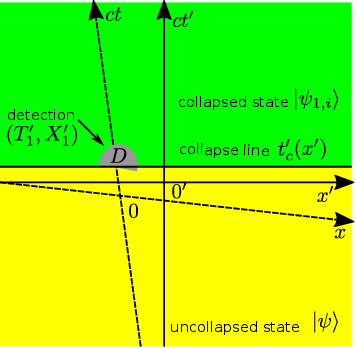}
\caption{Space-time diagram of evolution of a QS $\vert \psi \rangle$ in reference frame $S' \equiv(ct',x')$ undergoing measurement by detector $D$ at space-time point $(T_1',X_1')$. The detector $D$ is stationary in reference frame $S \equiv (ct,x)$. The evolution of the QS takes into account A\&A collapse model. After the measurement, QS $\vert \psi \rangle$ collapses into state $\vert \psi_{1,i} \rangle$ according to Eq.~(\ref{e1}). The evolution of the QS $\vert \psi \rangle$ is depicted in reference frame $S'$, which moves in positive direction of axis $x$. The boundary dividing the collapsed and uncollapsed state is marked as collapse line.}
\label{fig7} 
\end{figure}

Let the observer be stationary in reference frame $S'$, which is moving with velocity $v > 0$ in reference frame $S$. Then, the collapse line created by measurement of detector $D$ on QS $\vert \psi \rangle$ is instantaneous in the observer's reference frame $S'$, see Fig.~\ref{fig7}. 

\begin{figure}
\includegraphics[scale=1]{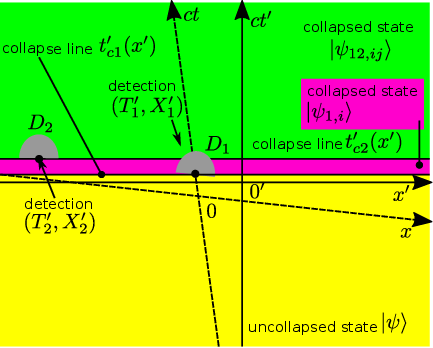}
\caption{Space-time diagram showing evolution of a QS $\vert \psi \rangle$ according to A\&A collapse model in reference frame $S' \equiv (ct',x')$. The quantum state $\vert \psi \rangle$ is measured by two detectors -- $D_1$ at space-time point $(T_1',X_1')$ and $D_2$ at space-time point $(T_2',X_2')$.  Detector $D_1$ and $D_2$ are at rest in reference frames $S\equiv(ct,x)$ and $S'$, respectively. The observer is associated with detector $D_2$. Reference frame $S'$ moves in positive direction of axis $x$. After measurement by detector $D_1$, the QS $\vert \psi \rangle$ collapses into state $\vert \psi_{1,i} \rangle$, according to Eq.~(\ref{e1}). After the second measurement by detector $D_2$ the state $\vert \psi_{1,i} \rangle$ collapses into state $\vert \psi_{12,ij} \rangle$.}
\label{fig8}
\end{figure}

In the same way as for MLF, H\&K and PLF models, measurement performed by two detectors $D_1$ and $D_2$ can be analyzed by means of A\&A model, see Fig.~\ref{fig8}. The detector $D_1$ ($D_2$) is assumed to be stationary in reference frame $S$ ($S'$). The reference frame $S'$ propagates with constant positive velocity $v$ in reference frame $S$. The observer is associated with detector $D_2$. In A\&A model, regardless of velocity of both detectors, the collapse lines will always be instantaneous in observer's reference frame ($S'$). Therefore, the measurements by detectors $D_1$ and $D_2$ will always be sequential and preserve correlations established in the initial state $\vert \psi \rangle$. With respect to collapse lines, A\&A model is equivalent to MLF model, when all detectors are stationary in observer's reference frame.

\section{Triggering a collapse of a QS prior to a measurement in MLF collapse model}
\begin{figure}
\includegraphics[scale=1]{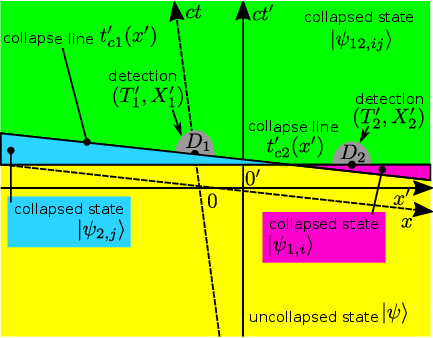}
\caption{Space-time scheme of evolution of a QS $\vert \psi \rangle$. The QS is subjected to two measurements by detectors $D_1$ and $D_2$. The detector $D_1$ ($D_2$) is in rest in reference frame $S$ ($S'$). The reference frame $S'$ moves in positive direction of axis $x$ with constant velocity. The detector $D_1$ and $D_2$ are taking measurements at events $(T_1',X_1')$ and $(T_2',X_2')$, respectively. The detectors are positioned to take measurements on already collapsed states $\vert \psi_{1,i} \rangle$ and $\vert \psi_{2,j} \rangle$. The utilized collapse model is MLF attached to a rest frame of a detector. The space-time area occupied by the initial state $\vert \psi \rangle$ has a yellow color. The space-time area occupied by state $\vert \psi_{1,i} \rangle$, which originates after projective measurement of detector $D_1$, has magenta color. Accordingly, QS $\vert \psi_{2,j} \rangle$ occupies a blue region. The regions are divided by collapse lines $t_{c1}'$ and $t_{c2}'$. Please do not be confused by a small rectangular area with a distinct color inside a larger one. Its intention is to highlight relation of its caption to the area that the caption belongs to.}
\label{fig9}
\end{figure}

The detector $D_2$ in Fig.~\ref{fig2} can be arranged in space $x'$, such that both detectors $D_1$ and $D_2$ perform a measurement on already collapsed states. This occurs, when detector $D_2$ is placed in blue location in Fig.~\ref{fig2}, where the state $\vert \psi \rangle$ is already collapsed to state $\vert \psi_{1,i} \rangle$. For space-time scheme of this situation, please see Fig.~\ref{fig9}. The detector $D_1$ (at rest in reference frame $S$) performs a measurement on state $\vert \psi_{2,j} \rangle$, which is given by a projective measurement of detector $D_2$ on state $\vert \psi_{1,i} \rangle$. This situation is similar to one, which always emerges when H\&K model is in use, see Fig.~\ref{fig6} and its discussion in Appendix~B. Due to our best knowledge, this situation has been briefly considered only by Zbinden~et~al.\cite{Zbinden2001experimental}, based on work of Suarez~\cite{Suarez1997does,Suarez1997relativistic}. Zbinden~et~al.\cite{Zbinden2001experimental} analyzed this situation by means of Suarez's theoretical model of relativistic measurement called ``Alternative description'' (AD). For more details, please read Sec.~I of this paper. This situation is treated in this appendix by means of MLF collapse model.

From Fig.~\ref{fig9} follows mathematical relations between collapsed and uncollapsed states
\begin{equation}
\begin{array}{rclrcl}
\vert \psi_{12,ij} \rangle &=& \hat{P}_{1,i} \vert \psi_{2,j} \rangle,& \vert \psi_{12,ij} \rangle &=& \hat{P}_{2,j} \vert \psi_{1,i} \rangle; \\[2mm]
\vert \psi_{1,i} \rangle &=& \hat{P}_{1,i} \vert \psi_{2,j} \rangle,&\quad \vert \psi_{2,j} \rangle &=& \hat{P}_{2,j} \vert \psi_{1,i} \rangle;\\[2mm]
\hat{P}_{1,i} &=& \vert \phi_{i} \rangle_1 {}_1\langle \phi_{i} \vert,&\quad \hat{P}_{2,j} &=& \vert \Psi_{j} \rangle_2 {}_2\langle \Psi_{j} \vert.
\end{array}
\label{eC1}
\end{equation}
Operators $\hat{P}_{1,i}$ and $\hat{P}_{2,j}$ are related to a projective measurements by detectors $D_1$ and $D_2$, respectively. After the measurement by detector $D_1$ ($D_2$), the state $\vert \psi_{2,j}\rangle$ ($\vert \psi_{1,i}\rangle$) collapses into state $\vert \psi_{1,i} \rangle_{}$ ($\vert \psi_{2,j} \rangle_{}$) according to definition of the projective measurement in Eq.~(\ref{e1}). 

From set of Eqs.~(\ref{eC1}) follows, that the states $\vert \psi_{12,ij} \rangle$, $\vert \psi_{1,i} \rangle$ and  $\vert \psi_{2,j} \rangle$ have to be all equal to
\begin{equation}
\vert \psi_{12,ij} \rangle = \vert \psi_{1,i} \rangle = \vert \psi_{2,j} \rangle = \vert \phi_{i} \rangle_1 \vert \Psi_{j} \rangle_2.
\end{equation}
Therefore, the magenta and blue regions in Fig.~\ref{fig9} should be attached a green color and be occupied by the already collapsed state $\vert \psi_{12,ij} \rangle$. In summary, in this situation detectors $D_1$ and $D_2$ perform a measurement on already collapsed state $\vert \psi_{12,ij} \rangle$. The initial state $\vert \psi \rangle$ collapses to final state $\vert \psi_{12,ij} \rangle$ in finite time before any measurement by either detector $D_1$ or $D_2$ takes place.

\bibliography{javurek.bib}

\begin{thebibliography}{32}%
\makeatletter
\providecommand \@ifxundefined [1]{%
 \@ifx{#1\undefined}
}%
\providecommand \@ifnum [1]{%
 \ifnum #1\expandafter \@firstoftwo
 \else \expandafter \@secondoftwo
 \fi
}%
\providecommand \@ifx [1]{%
 \ifx #1\expandafter \@firstoftwo
 \else \expandafter \@secondoftwo
 \fi
}%
\providecommand \natexlab [1]{#1}%
\providecommand \enquote  [1]{``#1''}%
\providecommand \bibnamefont  [1]{#1}%
\providecommand \bibfnamefont [1]{#1}%
\providecommand \citenamefont [1]{#1}%
\providecommand \href@noop [0]{\@secondoftwo}%
\providecommand \href [0]{\begingroup \@sanitize@url \@href}%
\providecommand \@href[1]{\@@startlink{#1}\@@href}%
\providecommand \@@href[1]{\endgroup#1\@@endlink}%
\providecommand \@sanitize@url [0]{\catcode `\\12\catcode `\$12\catcode
  `\&12\catcode `\#12\catcode `\^12\catcode `\_12\catcode `\%12\relax}%
\providecommand \@@startlink[1]{}%
\providecommand \@@endlink[0]{}%
\providecommand \url  [0]{\begingroup\@sanitize@url \@url }%
\providecommand \@url [1]{\endgroup\@href {#1}{\urlprefix }}%
\providecommand \urlprefix  [0]{URL }%
\providecommand \Eprint [0]{\href }%
\providecommand \doibase [0]{https://doi.org/}%
\providecommand \selectlanguage [0]{\@gobble}%
\providecommand \bibinfo  [0]{\@secondoftwo}%
\providecommand \bibfield  [0]{\@secondoftwo}%
\providecommand \translation [1]{[#1]}%
\providecommand \BibitemOpen [0]{}%
\providecommand \bibitemStop [0]{}%
\providecommand \bibitemNoStop [0]{.\EOS\space}%
\providecommand \EOS [0]{\spacefactor3000\relax}%
\providecommand \BibitemShut  [1]{\csname bibitem#1\endcsname}%
\let\auto@bib@innerbib\@empty
\bibitem [{\citenamefont {Susskind}\ and\ \citenamefont
  {Friedman}(2014)}]{Susskind2014quantum}%
  \BibitemOpen
  \bibfield  {author} {\bibinfo {author} {\bibfnamefont {L.}~\bibnamefont
  {Susskind}}\ and\ \bibinfo {author} {\bibfnamefont {A.}~\bibnamefont
  {Friedman}},\ }\href {https://books.google.cz/books?id=LX2-AQAAQBAJ} {\emph
  {\bibinfo {title} {Quantum Mechanics: The Theoretical Minimum}}}\ (\bibinfo
  {publisher} {Penguin Books Limited},\ \bibinfo {year} {2014})\BibitemShut
  {NoStop}%
\bibitem [{\citenamefont {Griffiths}(2004)}]{Griffiths2004introduction}%
  \BibitemOpen
  \bibfield  {author} {\bibinfo {author} {\bibfnamefont {D.~J.}\ \bibnamefont
  {Griffiths}},\ }\href
  {http://www.amazon.com/exec/obidos/redirect?tag=citeulike07-20\&path=ASIN/0131118927}
  {\emph {\bibinfo {title} {{Introduction to Quantum Mechanics (2nd
  Edition)}}}},\ \bibinfo {edition} {2nd}\ ed.\ (\bibinfo  {publisher} {Pearson
  Prentice Hall},\ \bibinfo {year} {2004})\BibitemShut {NoStop}%
\bibitem [{\citenamefont {Wechsler}(2021)}]{Wechsler2021quantum}%
  \BibitemOpen
  \bibfield  {author} {\bibinfo {author} {\bibfnamefont {S.~D.}\ \bibnamefont
  {Wechsler}},\ }\bibfield  {title} {\bibinfo {title} {The quantum mechanics
  needs the principle of wave-function collapse--but this principle shouldn't
  be misunderstood},\ }\href {https://doi.org/10.4236/jqis.2021.111004}
  {\bibfield  {journal} {\bibinfo  {journal} {Journal of Quantum Information
  Science}\ }\textbf {\bibinfo {volume} {11}},\ \bibinfo {pages} {42} (\bibinfo
  {year} {2021})}\BibitemShut {NoStop}%
\bibitem [{\citenamefont {Ohanian}(2017)}]{Ohanian2017}%
  \BibitemOpen
  \bibfield  {author} {\bibinfo {author} {\bibfnamefont {H.~C.}\ \bibnamefont
  {Ohanian}},\ }\bibfield  {title} {\bibinfo {title} {Collapse of probability
  distributions in relativistic spacetime},\ }\Eprint
  {https://arxiv.org/abs/1703.00309v1} {arXiv:1703.00309v1 [quant-ph]}
  (\bibinfo {year} {2017})\BibitemShut {NoStop}%
\bibitem [{\citenamefont {Bloch}(1967)}]{Bloch1967some}%
  \BibitemOpen
  \bibfield  {author} {\bibinfo {author} {\bibfnamefont {I.}~\bibnamefont
  {Bloch}},\ }\bibfield  {title} {\bibinfo {title} {Some relativistic oddities
  in the quantum theory of observation},\ }\href
  {https://doi.org/10.1103/PhysRev.156.1377} {\bibfield  {journal} {\bibinfo
  {journal} {Phys. Rev.}\ }\textbf {\bibinfo {volume} {156}},\ \bibinfo {pages}
  {1377} (\bibinfo {year} {1967})}\BibitemShut {NoStop}%
\bibitem [{\citenamefont {Hellwig}\ and\ \citenamefont
  {Kraus}(1970)}]{Hellwig1970formal}%
  \BibitemOpen
  \bibfield  {author} {\bibinfo {author} {\bibfnamefont {K.~E.}\ \bibnamefont
  {Hellwig}}\ and\ \bibinfo {author} {\bibfnamefont {K.}~\bibnamefont
  {Kraus}},\ }\bibfield  {title} {\bibinfo {title} {Formal description of
  measurements in local quantum field theory},\ }\href
  {https://doi.org/10.1103/PhysRevD.1.566} {\bibfield  {journal} {\bibinfo
  {journal} {Phys. Rev. D}\ }\textbf {\bibinfo {volume} {1}},\ \bibinfo {pages}
  {566} (\bibinfo {year} {1970})}\BibitemShut {NoStop}%
\bibitem [{\citenamefont {Aharonov}\ and\ \citenamefont
  {Albert}(1981)}]{Aharonov1981can}%
  \BibitemOpen
  \bibfield  {author} {\bibinfo {author} {\bibfnamefont {Y.}~\bibnamefont
  {Aharonov}}\ and\ \bibinfo {author} {\bibfnamefont {D.~Z.}\ \bibnamefont
  {Albert}},\ }\bibfield  {title} {\bibinfo {title} {Can we make sense out of
  the measurement process in relativistic quantum mechanics?},\ }\href
  {https://doi.org/10.1103/PhysRevD.24.359} {\bibfield  {journal} {\bibinfo
  {journal} {Phys. Rev. D}\ }\textbf {\bibinfo {volume} {24}},\ \bibinfo
  {pages} {359} (\bibinfo {year} {1981})}\BibitemShut {NoStop}%
\bibitem [{\citenamefont {Cohen}\ and\ \citenamefont
  {Hiley}(1995)}]{Cohen1995retrodiction}%
  \BibitemOpen
  \bibfield  {author} {\bibinfo {author} {\bibfnamefont {O.}~\bibnamefont
  {Cohen}}\ and\ \bibinfo {author} {\bibfnamefont {B.~J.}\ \bibnamefont
  {Hiley}},\ }\bibfield  {title} {\bibinfo {title} {Retrodiction in quantum
  mechanics, preferred lorentz frames, and nonlocal measurements},\ }\href
  {https://doi.org/10.1007/BF02057882} {\bibfield  {journal} {\bibinfo
  {journal} {Foundations of Physics}\ }\textbf {\bibinfo {volume} {25}},\
  \bibinfo {pages} {1669} (\bibinfo {year} {1995})}\BibitemShut {NoStop}%
\bibitem [{\citenamefont {Maldacena}\ and\ \citenamefont
  {Susskind}(2013)}]{Maldacena2013cool}%
  \BibitemOpen
  \bibfield  {author} {\bibinfo {author} {\bibfnamefont {J.}~\bibnamefont
  {Maldacena}}\ and\ \bibinfo {author} {\bibfnamefont {L.}~\bibnamefont
  {Susskind}},\ }\bibfield  {title} {\bibinfo {title} {Cool horizons for
  entangled black holes},\ }\href {https://doi.org/10.1002/prop.201300020}
  {\bibfield  {journal} {\bibinfo  {journal} {Fortschritte der Physik}\
  }\textbf {\bibinfo {volume} {61}},\ \bibinfo {pages} {781} (\bibinfo {year}
  {2013})}\BibitemShut {NoStop}%
\bibitem [{\citenamefont {Susskind}(2016)}]{Susskind2016copenhagen}%
  \BibitemOpen
  \bibfield  {author} {\bibinfo {author} {\bibfnamefont {L.}~\bibnamefont
  {Susskind}},\ }\bibfield  {title} {\bibinfo {title} {Copenhagen vs {Everett},
  teleportation, and {ER}={EPR}},\ }\href
  {https://doi.org/10.1002/prop.201600036} {\bibfield  {journal} {\bibinfo
  {journal} {Fortschritte der Physik}\ }\textbf {\bibinfo {volume} {64}},\
  \bibinfo {pages} {551} (\bibinfo {year} {2016})}\BibitemShut {NoStop}%
\bibitem [{Note1()}]{Note1}%
  \BibitemOpen
  \bibinfo {note} {Observer does not have to posses a detector and perform a
  measurement in this context.}\BibitemShut {Stop}%
\bibitem [{\citenamefont {Suarez}\ and\ \citenamefont
  {Scarani}(1997)}]{Suarez1997does}%
  \BibitemOpen
  \bibfield  {author} {\bibinfo {author} {\bibfnamefont {A.}~\bibnamefont
  {Suarez}}\ and\ \bibinfo {author} {\bibfnamefont {V.}~\bibnamefont
  {Scarani}},\ }\bibfield  {title} {\bibinfo {title} {Does entanglement depend
  on the timing of the impacts at the beam-splitters?},\ }\href
  {https://doi.org/https://doi.org/10.1016/S0375-9601(97)00346-0} {\bibfield
  {journal} {\bibinfo  {journal} {Physics Letters A}\ }\textbf {\bibinfo
  {volume} {232}},\ \bibinfo {pages} {9} (\bibinfo {year} {1997})}\BibitemShut
  {NoStop}%
\bibitem [{\citenamefont {Zbinden}\ \emph {et~al.}(2001)\citenamefont
  {Zbinden}, \citenamefont {Brendel}, \citenamefont {Gisin},\ and\
  \citenamefont {Tittel}}]{Zbinden2001experimental}%
  \BibitemOpen
  \bibfield  {author} {\bibinfo {author} {\bibfnamefont {H.}~\bibnamefont
  {Zbinden}}, \bibinfo {author} {\bibfnamefont {J.}~\bibnamefont {Brendel}},
  \bibinfo {author} {\bibfnamefont {N.}~\bibnamefont {Gisin}},\ and\ \bibinfo
  {author} {\bibfnamefont {W.}~\bibnamefont {Tittel}},\ }\bibfield  {title}
  {\bibinfo {title} {Experimental test of nonlocal quantum correlation in
  relativistic configurations},\ }\href
  {https://doi.org/10.1103/PhysRevA.63.022111} {\bibfield  {journal} {\bibinfo
  {journal} {Phys. Rev. A}\ }\textbf {\bibinfo {volume} {63}},\ \bibinfo
  {pages} {022111} (\bibinfo {year} {2001})}\BibitemShut {NoStop}%
\bibitem [{\citenamefont {Pater}\ \emph {et~al.}(2021)\citenamefont {Pater},
  \citenamefont {Atteya},\ and\ \citenamefont {Tariq}}]{Pater2021temporal}%
  \BibitemOpen
  \bibfield  {author} {\bibinfo {author} {\bibfnamefont {C.}~\bibnamefont
  {Pater}}, \bibinfo {author} {\bibfnamefont {D.}~\bibnamefont {Atteya}},\ and\
  \bibinfo {author} {\bibfnamefont {S.}~\bibnamefont {Tariq}},\ }\bibfield
  {title} {\bibinfo {title} {Temporal ordering of the wavefunction collapse in
  relativity},\ }\bibfield  {journal} {\bibinfo  {journal} {Researchgate}\
  }\href {https://doi.org/10.13140/RG.2.2.21425.30568}
  {10.13140/RG.2.2.21425.30568} (\bibinfo {year} {2021})\BibitemShut {NoStop}%
\bibitem [{\citenamefont {Salart}\ \emph {et~al.}(2008)\citenamefont {Salart},
  \citenamefont {Baas}, \citenamefont {Branciard}, \citenamefont {Gisin},\ and\
  \citenamefont {Zbinden}}]{Salart2008}%
  \BibitemOpen
  \bibfield  {author} {\bibinfo {author} {\bibfnamefont {D.}~\bibnamefont
  {Salart}}, \bibinfo {author} {\bibfnamefont {A.}~\bibnamefont {Baas}},
  \bibinfo {author} {\bibfnamefont {C.}~\bibnamefont {Branciard}}, \bibinfo
  {author} {\bibfnamefont {N.}~\bibnamefont {Gisin}},\ and\ \bibinfo {author}
  {\bibfnamefont {H.}~\bibnamefont {Zbinden}},\ }\bibfield  {title} {\bibinfo
  {title} {Testing the speed of `spooky action at a distance'},\ }\href
  {https://doi.org/10.1038/nature07121} {\bibfield  {journal} {\bibinfo
  {journal} {Nature}\ }\textbf {\bibinfo {volume} {454}},\ \bibinfo {pages}
  {861} (\bibinfo {year} {2008})}\BibitemShut {NoStop}%
\bibitem [{\citenamefont {Scarani}\ \emph {et~al.}(2000)\citenamefont
  {Scarani}, \citenamefont {Tittel}, \citenamefont {Zbinden},\ and\
  \citenamefont {Gisin}}]{Scarani2000speed}%
  \BibitemOpen
  \bibfield  {author} {\bibinfo {author} {\bibfnamefont {V.}~\bibnamefont
  {Scarani}}, \bibinfo {author} {\bibfnamefont {W.}~\bibnamefont {Tittel}},
  \bibinfo {author} {\bibfnamefont {H.}~\bibnamefont {Zbinden}},\ and\ \bibinfo
  {author} {\bibfnamefont {N.}~\bibnamefont {Gisin}},\ }\bibfield  {title}
  {\bibinfo {title} {The speed of quantum information and the preferred frame:
  analysis of experimental data},\ }\href
  {https://doi.org/https://doi.org/10.1016/S0375-9601(00)00609-5} {\bibfield
  {journal} {\bibinfo  {journal} {Physics Letters A}\ }\textbf {\bibinfo
  {volume} {276}},\ \bibinfo {pages} {1} (\bibinfo {year} {2000})}\BibitemShut
  {NoStop}%
\bibitem [{\citenamefont {Weihs}\ \emph {et~al.}(1998)\citenamefont {Weihs},
  \citenamefont {Jennewein}, \citenamefont {Simon}, \citenamefont
  {Weinfurter},\ and\ \citenamefont {Zeilinger}}]{Weihs1998violation}%
  \BibitemOpen
  \bibfield  {author} {\bibinfo {author} {\bibfnamefont {G.}~\bibnamefont
  {Weihs}}, \bibinfo {author} {\bibfnamefont {T.}~\bibnamefont {Jennewein}},
  \bibinfo {author} {\bibfnamefont {C.}~\bibnamefont {Simon}}, \bibinfo
  {author} {\bibfnamefont {H.}~\bibnamefont {Weinfurter}},\ and\ \bibinfo
  {author} {\bibfnamefont {A.}~\bibnamefont {Zeilinger}},\ }\bibfield  {title}
  {\bibinfo {title} {Violation of bell's inequality under strict einstein
  locality conditions},\ }\href {https://doi.org/10.1103/PhysRevLett.81.5039}
  {\bibfield  {journal} {\bibinfo  {journal} {Phys. Rev. Lett.}\ }\textbf
  {\bibinfo {volume} {81}},\ \bibinfo {pages} {5039} (\bibinfo {year}
  {1998})}\BibitemShut {NoStop}%
\bibitem [{\citenamefont {Suarez}(2000)}]{Suarez2000preferred}%
  \BibitemOpen
  \bibfield  {author} {\bibinfo {author} {\bibfnamefont {A.}~\bibnamefont
  {Suarez}},\ }\href {https://doi.org/10.48550/ARXIV.QUANT-PH/0006053}
  {\bibinfo {title} {Preferred frame versus multisimultaneity: meaning and
  relevance of a forthcoming experiment}} (\bibinfo {year} {2000}),\ \Eprint
  {https://arxiv.org/abs/0006053v1} {arXiv:0006053v1 [quant-ph]} \BibitemShut
  {NoStop}%
\bibitem [{Note2()}]{Note2}%
  \BibitemOpen
  \bibinfo {note} {According to Cohen and Hiley \cite {Cohen1995retrodiction},
  assuming independent existence of a QS regardless of a measurement implies,
  that standard interpretation of quantum mechanics is being used.}\BibitemShut
  {Stop}%
\bibitem [{\citenamefont {Garrisi}\ \emph {et~al.}(2019)\citenamefont
  {Garrisi}, \citenamefont {Massara}, \citenamefont {Zambianchi}, \citenamefont
  {Galli}, \citenamefont {Bajoni}, \citenamefont {Rimini},\ and\ \citenamefont
  {Nicrosini}}]{Garrisi2019experimental}%
  \BibitemOpen
  \bibfield  {author} {\bibinfo {author} {\bibfnamefont {F.}~\bibnamefont
  {Garrisi}}, \bibinfo {author} {\bibfnamefont {M.}~\bibnamefont {Massara}},
  \bibinfo {author} {\bibfnamefont {A.}~\bibnamefont {Zambianchi}}, \bibinfo
  {author} {\bibfnamefont {M.}~\bibnamefont {Galli}}, \bibinfo {author}
  {\bibfnamefont {D.}~\bibnamefont {Bajoni}}, \bibinfo {author} {\bibfnamefont
  {A.}~\bibnamefont {Rimini}},\ and\ \bibinfo {author} {\bibfnamefont
  {O.}~\bibnamefont {Nicrosini}},\ }\bibfield  {title} {\bibinfo {title}
  {Experimental test of the collapse time of a delocalized photon state},\
  }\href {https://doi.org/10.1038/s41598-019-48387-8} {\bibfield  {journal}
  {\bibinfo  {journal} {Scientific Reports}\ }\textbf {\bibinfo {volume} {9}}
  (\bibinfo {year} {2019})}\BibitemShut {NoStop}%
\bibitem [{\citenamefont {Moreno}\ and\ \citenamefont
  {Parisio}(2013)}]{Moreno2013looking}%
  \BibitemOpen
  \bibfield  {author} {\bibinfo {author} {\bibfnamefont {M.}~\bibnamefont
  {Moreno}}\ and\ \bibinfo {author} {\bibfnamefont {F.}~\bibnamefont
  {Parisio}},\ }\bibfield  {title} {\bibinfo {title} {Looking into the collapse
  of quantum states with entangled photons},\ }\Eprint
  {https://arxiv.org/abs/1303.2972v1} {arXiv:1303.2972v1 [quant-ph]}  (\bibinfo
  {year} {2013})\BibitemShut {NoStop}%
\bibitem [{Note3()}]{Note3}%
  \BibitemOpen
  \bibinfo {note} {We can find one observer (moving with fixed pre-defined
  velocity), in which reference frame the time of detection of detector $D_1$
  precedes detection by detector $D_2$. Example of this reference frame is $S'$
  (see Fig.~\ref {fig2}). On the other hand, we can find another observer
  (moving with different fixed velocity), in which reference frame detection of
  detector $D_2$ precedes detection of detector $D_1$. Example of this
  reference frame is $S$ (see Fig.~\ref {fig2}).}\BibitemShut {Stop}%
\bibitem [{Note4()}]{Note4}%
  \BibitemOpen
  \bibinfo {note} {In this case, the time dilatation caused by general
  relativity effects would have to be considered as well.}\BibitemShut {Stop}%
\bibitem [{Note5()}]{Note5}%
  \BibitemOpen
  \bibinfo {note} {In practice, the photon pair can be generated by a
  femto-second pump-beam pulse. It generates photon pair with pulse lengths in
  order of hundreds of femtoseconds \cite {Perina1999dispersion}. Therefore,
  the pulse lengths of photons in the pair are much smaller than the time
  required for detection and storage $\Delta t_s \sim 0.1$~ns. This guarantees
  successful detection of both photons in the pair at required times $T_1$ and
  $T_2$ with negligible error.}\BibitemShut {Stop}%
\bibitem [{\citenamefont {Suarez}(1997)}]{Suarez1997relativistic}%
  \BibitemOpen
  \bibfield  {author} {\bibinfo {author} {\bibfnamefont {A.}~\bibnamefont
  {Suarez}},\ }\bibfield  {title} {\bibinfo {title} {Relativistic nonlocality
  in an experiment with 2 non-before impacts},\ }\href
  {https://doi.org/https://doi.org/10.1016/S0375-9601(97)00804-9} {\bibfield
  {journal} {\bibinfo  {journal} {Physics Letters A}\ }\textbf {\bibinfo
  {volume} {236}},\ \bibinfo {pages} {383} (\bibinfo {year}
  {1997})}\BibitemShut {NoStop}%
\bibitem [{\citenamefont {Arkhipov}\ \emph {et~al.}(2015)\citenamefont
  {Arkhipov}, \citenamefont {Pe\ifmmode~\check{r}\else \v{r}\fi{}ina},
  \citenamefont {Pe\ifmmode~\check{r}\else \v{r}\fi{}ina},\ and\ \citenamefont
  {Miranowicz}}]{Arkhipov2015comparative}%
  \BibitemOpen
  \bibfield  {author} {\bibinfo {author} {\bibfnamefont {I.~I.}\ \bibnamefont
  {Arkhipov}}, \bibinfo {author} {\bibfnamefont {J.}~\bibnamefont
  {Pe\ifmmode~\check{r}\else \v{r}\fi{}ina}}, \bibinfo {author} {\bibfnamefont
  {J.}~\bibnamefont {Pe\ifmmode~\check{r}\else \v{r}\fi{}ina}},\ and\ \bibinfo
  {author} {\bibfnamefont {A.}~\bibnamefont {Miranowicz}},\ }\bibfield  {title}
  {\bibinfo {title} {Comparative study of nonclassicality, entanglement, and
  dimensionality of multimode noisy twin beams},\ }\href
  {https://doi.org/10.1103/PhysRevA.91.033837} {\bibfield  {journal} {\bibinfo
  {journal} {Phys. Rev. A}\ }\textbf {\bibinfo {volume} {91}},\ \bibinfo
  {pages} {033837} (\bibinfo {year} {2015})}\BibitemShut {NoStop}%
\bibitem [{\citenamefont {Pe\ifmmode~\check{r}\else
  \v{r}\fi{}ina}(2016)}]{Perina2016spatial}%
  \BibitemOpen
  \bibfield  {author} {\bibinfo {author} {\bibfnamefont {J.}~\bibnamefont
  {Pe\ifmmode~\check{r}\else \v{r}\fi{}ina}},\ }\bibfield  {title} {\bibinfo
  {title} {Spatial, spectral, and temporal coherence of ultraintense twin
  beams},\ }\href {https://doi.org/10.1103/PhysRevA.93.013852} {\bibfield
  {journal} {\bibinfo  {journal} {Phys. Rev. A}\ }\textbf {\bibinfo {volume}
  {93}},\ \bibinfo {pages} {013852} (\bibinfo {year} {2016})}\BibitemShut
  {NoStop}%
\bibitem [{\citenamefont {Reiserer}\ \emph {et~al.}(2013)\citenamefont
  {Reiserer}, \citenamefont {Ritter},\ and\ \citenamefont
  {Rempe}}]{Reiserer2013nondestructive}%
  \BibitemOpen
  \bibfield  {author} {\bibinfo {author} {\bibfnamefont {A.}~\bibnamefont
  {Reiserer}}, \bibinfo {author} {\bibfnamefont {S.}~\bibnamefont {Ritter}},\
  and\ \bibinfo {author} {\bibfnamefont {G.}~\bibnamefont {Rempe}},\ }\bibfield
   {title} {\bibinfo {title} {Nondestructive detection of an optical photon},\
  }\href {https://doi.org/10.1126/science.1246164} {\bibfield  {journal}
  {\bibinfo  {journal} {Science}\ }\textbf {\bibinfo {volume} {342}},\ \bibinfo
  {pages} {1349} (\bibinfo {year} {2013})},\ \Eprint
  {https://arxiv.org/abs/https://www.science.org/doi/pdf/10.1126/science.1246164}
  {https://www.science.org/doi/pdf/10.1126/science.1246164} \BibitemShut
  {NoStop}%
\bibitem [{\citenamefont {Niemietz}\ \emph {et~al.}(2021)\citenamefont
  {Niemietz}, \citenamefont {Farrera}, \citenamefont {Lagenfeld},\ and\
  \citenamefont {Gerhard}}]{Niemietz2021nondestructive}%
  \BibitemOpen
  \bibfield  {author} {\bibinfo {author} {\bibfnamefont {D.}~\bibnamefont
  {Niemietz}}, \bibinfo {author} {\bibfnamefont {P.}~\bibnamefont {Farrera}},
  \bibinfo {author} {\bibfnamefont {S.}~\bibnamefont {Lagenfeld}},\ and\
  \bibinfo {author} {\bibfnamefont {R.}~\bibnamefont {Gerhard}},\ }\bibfield
  {title} {\bibinfo {title} {Nondestructive detection of photonic qubits},\
  }\href {https://doi.org/10.1038/s41586-021-03290-z} {\bibfield  {journal}
  {\bibinfo  {journal} {Nature}\ }\textbf {\bibinfo {volume} {591}},\ \bibinfo
  {pages} {570} (\bibinfo {year} {2021})}\BibitemShut {NoStop}%
\bibitem [{Note6()}]{Note6}%
  \BibitemOpen
  \bibinfo {note} {To prove this statement, normalization condition for the
  probability density $p$, $\DOTSI \protect \MultiIntegral {2}_0^\infty d
  \omega _s d \omega _i\protect \, p (\omega _s,\omega _i) = 1$ has to be
  used.}\BibitemShut {Stop}%
\bibitem [{Note7()}]{Note7}%
  \BibitemOpen
  \bibinfo {note} {In~\cite {Hellwig1970formal}, see Fig.~3 together with its
  description in paragraph above Eq.~(3) and definition in
  Eq.~(2).}\BibitemShut {Stop}%
\bibitem [{\citenamefont {Pe\ifmmode~\check{r}\else \v{r}\fi{}ina}\ \emph
  {et~al.}(1999)\citenamefont {Pe\ifmmode~\check{r}\else \v{r}\fi{}ina},
  \citenamefont {Sergienko}, \citenamefont {Jost}, \citenamefont {Saleh},\ and\
  \citenamefont {Teich}}]{Perina1999dispersion}%
  \BibitemOpen
  \bibfield  {author} {\bibinfo {author} {\bibfnamefont {J.}~\bibnamefont
  {Pe\ifmmode~\check{r}\else \v{r}\fi{}ina}}, \bibinfo {author} {\bibfnamefont
  {A.~V.}\ \bibnamefont {Sergienko}}, \bibinfo {author} {\bibfnamefont {B.~M.}\
  \bibnamefont {Jost}}, \bibinfo {author} {\bibfnamefont {B.~E.~A.}\
  \bibnamefont {Saleh}},\ and\ \bibinfo {author} {\bibfnamefont {M.~C.}\
  \bibnamefont {Teich}},\ }\bibfield  {title} {\bibinfo {title} {Dispersion in
  femtosecond entangled two-photon interference},\ }\href
  {https://doi.org/10.1103/PhysRevA.59.2359} {\bibfield  {journal} {\bibinfo
  {journal} {Phys. Rev. A}\ }\textbf {\bibinfo {volume} {59}},\ \bibinfo
  {pages} {2359} (\bibinfo {year} {1999})}\BibitemShut {NoStop}%
\end{thebibliography}%

\end{document}